%% file: main.tex
\documentclass[sigconf]{acmart}
\renewcommand\footnotetextcopyrightpermission[1]{} 
\usepackage[utf8]{inputenc}
\usepackage{natbib}
\usepackage{graphicx}
\usepackage[caption=false]{subfig}
\usepackage{multirow}
\usepackage{multicol}
\usepackage{array}
\usepackage{xcolor}
\usepackage{float}
\usepackage[normalem]{ulem}
\usepackage{listings}
\usepackage{makecell}
\usepackage{hyperref}
\usepackage{xspace}

\newcommand{\dcdb}{DCDB\xspace}
\newcommand{\daf}{Wintermute\xspace}
\newcommand{\sng}{SuperMUC-NG\xspace}
\newcommand{\dpp}{DEEP-EST\xspace}
\newcommand{\lrz}{LRZ\xspace}
\newcommand{\jul}{JSC\xspace}
\newcommand{\persyst}{Persyst\xspace}
\newcommand{\csmethod}{CS\xspace}

\newcommand{\CM}{CM\xspace}
\newcommand{\ESB}{ESB\xspace}
\newcommand{\DAM}{DAM\xspace}

\title[Operational Data Analytics in Practice]{Operational Data Analytics in Practice: Experiences from Design to Deployment in Production HPC Environments}

\author{Alessio Netti}
\email{alessio.netti@lrz.de}
\affiliation{Leibniz Supercomputing Centre}
\affiliation{Technical University of Munich}
\author{Michael Ott}
\email{michael.ott@lrz.de}
\affiliation{Leibniz Supercomputing Centre}
\author{Carla Guillen}
\email{carla.guillen@lrz.de}
\affiliation{Leibniz Supercomputing Centre}
\author{Daniele Tafani}
\email{daniele.tafani@fujitsu.com}
\affiliation{Fujitsu Enabling Software Technology}
\author{Martin Schulz}
\email{schulzm@in.tum.de}
\affiliation{Technical University of Munich}
\affiliation{Leibniz Supercomputing Centre}

\acmConference[]{Preliminary version}{June 2021}{Munich, Germany}

\begin{document}

\begin{abstract}

As HPC systems grow in complexity, efficient and manageable operation is increasingly critical. Many centers are thus starting to explore the use of \emph{Operational Data Analytics} (ODA) techniques, which extract knowledge from massive amounts of monitoring data and use it for control and visualization purposes. As ODA is a multi-faceted problem, much effort has gone into researching its separate aspects: however, accounts of production ODA experiences are still hard to come across.

In this work we aim to bridge the gap between ODA research and production use by presenting our experiences with ODA in production, involving in particular the control of cooling infrastructures and visualization of job data on two HPC systems. We cover the entire development process, from design to deployment, highlighting our insights in an effort to drive the community forward. We rely on open-source tools, which make for a generic ODA framework suitable for most scenarios.
\end{abstract}


\keywords{High-Performance Computing, Monitoring, Operational Data Analytics, System Control, Data Mining}

\maketitle
\renewcommand{\shortauthors}{Netti et al.}

\input{sections/Introduction}
\input{sections/Design}
\input{sections/SNGDeployment}
\input{sections/DPDeployment}
\input{sections/LessonsLearned}
\input{sections/Conclusions}


\bibliographystyle{ACM-Reference-Format}
\bibliography{main}

\end{document}

%% file: sections/Introduction.tex
\section{Introduction}
\label{section:introduction}

Computation lies at the foundation of modern industry: large-scale \emph{High-Performance Computing} (HPC) systems drive scientific research through experimentation, while commercial data centers provide fundamental services for most businesses. As the demand for computational capacity grows, with the HPC community close to the \emph{exascale} goal, efficiency and sustainability of next-generation data centers has become the central theme. Modern HPC systems are massive in scale and complexity, comprising thousands of compute nodes and potentially tens of thousands of CPUs and GPUs, whose performance is subject to significant manufacturing variability~\cite{inadomi2015analyzing}; this complexity extends to the infrastructure (e.g., for cooling) and software stack levels, making production operation by itself a non-trivial task. Further, operating large-scale systems is becoming increasingly prohibitive from a cost standpoint, with excessive component failure rates~\cite{cappello2014toward} and energy consumption~\cite{villa2014scaling}. 

The advancement of hardware technologies must thus be coupled with improved usage of system resources. The \emph{Operational Data Analytics}~\cite{Bourassa:2019:ODA:3339186.3339210} (ODA) field is key in this endeavour, by providing techniques that are able to derive actionable knowledge from the massive amounts of sensor data produced by HPC systems. Due to the broadness of ODA, we introduce a classification with two high-level categories: with \emph{ODA for Visualization} (ODAV), ODA-derived information is visualized by system administrators to assist them in daily operations, whereas with \emph{ODA for Control} (ODAC) it is translated by the ODA system itself into new settings for system knobs. A typical example of ODAV is the visualization of aggregated user job metrics~\cite{Guillen2014, eitzinger2019clustercockpit} for performance characterization - it should be noted that ODAV implies the presence of one or more processing steps, and we do not classify the simple visualization of raw monitoring data under this category. On the other hand, ODAC embraces a much wider set of applications: these range from automatic CPU frequency optimization~\cite{eastep2017global}, to fault detection for improved reliability~\cite{tuncer2018online}, and from application classification for scheduling purposes~\cite{wyatt2018prionn} to the tuning of cooling infrastructures~\cite{jiang2019fine}. Most ODA techniques are designed to operate \emph{online} (i.e., in real time) and optimize system operations proactively.

However, as ODA is a multi-faceted problem comprising different stages, each with its own challenges and constraints, its implementation, deployment and use in production systems is in itself non-trivial: sensor data must be acquired using a scalable monitoring infrastructure, continuously covering the entire system; the data must then be processed into a useful representation via techniques that have to be lightweight as well as portable, before it can be finally leveraged by ODAV or ODAC algorithms, which by themselves must be lightweight, robust and able to extract insights from the collected data. Each of these individual steps has been explored thoroughly in the ODA research field, but there is currently a severe lack of end-to-end experiences, from design down to maintenance, covering all aspects of the ODA pipeline and providing the necessary insights and solutions to propel forward the capillary adoption of ODA in production data center environments. It has been observed in the literature, in fact, that most HPC centers rely on insular ODA solutions tackling only specific aspects of the problem~\cite{ott2020global}, with no clear applicability to other domains.

\paragraph{Contributions}

With this paper we aim to clear the haze around the use of ODA techniques in data center environments, by presenting our own ODAV and ODAC design, deployment and management experiences on several large-scale production HPC systems. We use a tightly-integrated pipeline of tools and techniques developed over the years, following a strict set of requirements - these compose an end-to-end, open-source framework that is capable of executing most ODA use cases and that is free for use by the community. Our experiences cover online visualization of job data (ODAV) on the large-scale \emph{\sng} HPC system at the \emph{Leibniz Supercomputing Centre} (LRZ), as well as proactive control of the direct-liquid cooling infrastructure (ODAC) of the modular \emph{\dpp} HPC system at the \emph{Juelich Supercomputing Centre} (JSC): we first cover the challenges and respective proposed solutions for each experience, then proceed with the design of the underlying ODA infrastructure and its deployment, and conclude with its evaluation in production operation. We finally discuss the complexity factors stemming from our experiences and propose action items for the community, particularly regarding the long-term maintainability of ODA. This way, we aim to provide a point of reference to ODA researchers and system administrators alike.

\paragraph{Organization}

The paper is structured as follows. In Section~\ref{section:design} we review related work and introduce the tools used for our deployments. In Section~\ref{section:sngdeployment} we then present our ODAV experience on \sng, and in Section~\ref{section:dpdeployment} our ODAC deployment on the \dpp system. In Section~\ref{section:lessonslearned} we discuss the main lessons learned based on our experiences, and in Section~\ref{section:conclusions} we conclude the paper. 

%% file: sections/Design.tex
\section{State of the Art and System Design}
\label{section:design}

\begin{figure}[t]
\centering
\includegraphics[width=0.45\textwidth,trim={0 0 0 0}, clip=true]{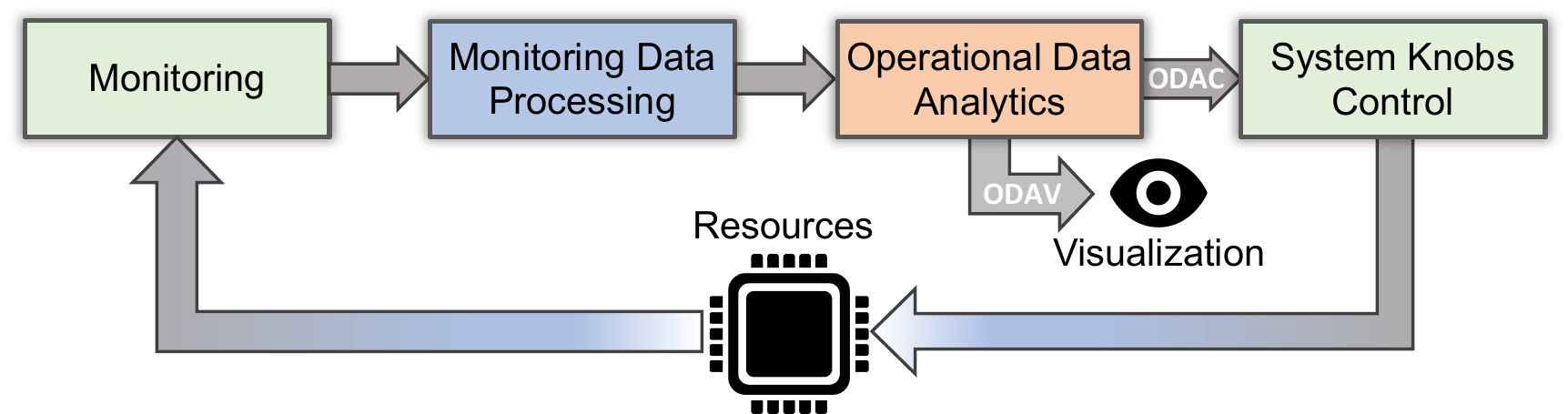}
\caption{The stages comprising a generic ODA pipeline.}
\label{design:dapipeline}
\end{figure}

Based on an extensive survey of experimental techniques proposed in the literature~\cite{conficoni2015energy, jiang2019fine, grant2015overtime,imes2018energy,sirbu2016power, galleguillos2017data,naghshnejad2018adaptive,emeras2015evalix, ates2018taxonomist,gallo2015analysis,wyatt2018prionn,mckenna2016machine, tuncer2018online,shaykhislamov2018approach, wang2017modular} and in light of our long-term experiences at \lrz~\cite{netti2019dcdb, netti2019wintermute, netti2020correlation}, we derive a generic formulation for ODA - namely, we identify the main functional steps composing it. These are summarized in Figure~\ref{design:dapipeline}: the first step consists in \emph{Monitoring} of system resources by collecting sensor or log data; this is followed by \emph{Monitoring Data Processing}, transforming the raw monitoring data into a polished representation that can be comprehended by ODA techniques - aggregation and dimensionality reduction, for example, are two common approaches to solve this task. In some cases, this step is optional. The actual \emph{Operational Data Analytics} step comes next: here, a model of a certain kind (e.g., machine learning) is applied to the data representations computed by the previous step, extracting knowledge that can be used to improve a system's operation. This could consist, for example, of a health diagnosis for a compute node for failure detection, or a prediction of a system's workload in the next few minutes for dynamic resource management. In the case of ODAV, the output of the ODA step is visualized immediately, aiding users and operators alike in their tasks. In the case of ODAC, on the other hand, it is further propagated to a \emph{System Knobs Control} step, which applies the ODA process's output as a new setting for a certain system knob using a specific interface (e.g., CPU frequency via \emph{SysFS}). This new setting affects the operation of the resources being monitored, establishing a feedback loop. Based on this classification, we now review the main accounts of ODA techniques that have been effectively implemented and deployed for production use, following with a list of operational requirements driving ODA processes, and with an overview of the tools we use for our deployments.

\subsection{State of the Art}

A wide variety of software solutions falling under our definition of ODA are available in the literature, aiming to cover the different aspects of an HPC system's operation~\cite{netti2019wintermute}. The vast majority of these techniques have never been employed in a production context, and thus their practical applicability is not proven: on the other hand, a small subset of tools are indeed known to have been used in production environments, but are either tailored for individual use cases or lack any report of long-term operation. Among these we find the \emph{Global Extensible Open Power Manager} (GEOPM)~\cite{eastep2017global} and the \emph{Energy-Aware Runtime} (EAR)~\cite{corbalan2019ear}, which are ODAC frameworks for CPU frequency tuning. Further, both the \emph{Lightweight Distributed Metric Service} (LDMS)~\cite{agelastos2014lightweight, izadpanah2018integrating} and the \emph{Examon}~\cite{beneventi2017continuous} monitoring frameworks are ODA-capable. Aside from open-source solutions, commercial ones, such as \emph{Nagios}~\cite{barth2008nagios}, \emph{Splunk}\footnote{\url{https://www.splunk.com}} or \emph{Icinga}\footnote{\url{https://www.icinga.com}}, are also common in HPC. While these support a certain degree of ODA, they are mostly alert-oriented and designed for loosely-coupled data centers. As such, they focus on infrastructure-level data, and complex ODA use cases (e.g., ODAC) are not supported.

Accounts of long-term production experiences, however, are currently still rare and mostly cover ODAV deployments. Brandt et al.~\cite{brandt2016large} discuss their 2-year LDMS deployment on \emph{Blue Waters}, a large-scale HPC system with more than 27,000 nodes. The authors describe a series of technical challenges related to reliability, overhead and data consistency, as well as dive into system-specific issues, such as clock skew effects. Similarly, Ahlgren et al.~\cite{ahlgren2018large} collect a series of experiences associated with monitoring and its main usage scenarios from several HPC centers. Here, the authors highlight the fact that most data centers rely on similar data sources (e.g., CPU performance counters), but at the same time employ highly different collection and storage solutions, either due to a lack of standard solutions or due to vendor constraints. Both works cover only monitoring and do not delve further into ODA issues. The term ODA in the HPC context is used for the first time by Bourassa et al.~\cite{Bourassa:2019:ODA:3339186.3339210} and Bautista et al.~\cite{bautista2019collecting}: these works describe several cases of successful infrastructure optimization and future system planning through visual inspection of data collected by the \emph{Operations Monitoring and Notification Infrastructure} (OMNI) framework at the \emph{National Energy Research Scientific Computing} (NERSC) center. Due to their visual nature, these works are ODAV case studies. Some works cover ODAC experiences for specific purposes: Auweter et al.~\cite{auweter2014case} discuss their use of the \emph{LoadLeveler} framework for CPU frequency tuning on the \emph{SuperMUC} HPC system at the \emph{Leibniz Supercomputing Centre} (LRZ), leading to 6\% yearly energy cost savings, while Jha et al.~\cite{jha2020live} describe their 2-year use of the \emph{Kaleidoscope} tool on Blue Waters for live failure detection.

The monitoring data processing aspect is rarely dealt with in the literature, with the exception of few works that treat specific use cases with ad-hoc techniques~\cite{tuncer2018online, bodik2010fingerprinting, lan2010toward, laguna2013automatic, hui2018comprehensive}. In general, the works listed so far highlight a key issue in the current use of ODA in data centers: there is a lack of holistic frameworks that encompass the totality of monitoring data sources and ODA use cases, from the infrastructure down to user applications, with their varying requirements and time scales. This often leads to multiple tools being used, in turn resulting in complex and fragmented software stacks and in a wide disarray of monitoring data that is difficult to use effectively~\cite{gimenez2017scrubjay}. This statement is confirmed in a survey conducted by the \emph{Energy Efficient HPC Working Group} (EEHPCWG) in 2019~\cite{ott2020global} regarding the use of ODA in several HPC centers: most sites employ varying sets of monitoring, storage and analysis solutions that either rely on in-house systems, or on commercial products that are not tailored for data center monitoring, thus restricting administrators to simple ODAV visual inspection.

\subsection{Requirements and Tools}

The single steps of ODA entail different operational requirements for working effectively in production environments. As these have been explored separately in previous work~\cite{netti2019dcdb, netti2019wintermute, netti2020correlation}, here we focus on the overarching requirements of an ODA pipeline as a whole: 

\begin{itemize}
    \item \textbf{Holism}: large-scale installations producing up to millions of heterogeneous metrics require the ability to collect, expose, process and finally use data in a uniform way, regardless of its source. This can be accomplished, for example, by using abstract interfaces and homogeneous data formats.
    \item \textbf{Scalability}: extreme volumes of data to collect and process are common in large-scale deployments. Hence, the pipeline must be distributed to support this kind of load.
    \item \textbf{Footprint}: interference on user applications, resource usage and daily operations in general must be minimal in order for an ODA pipeline to be cost-effective.
    \item \textbf{Modularity}: the ability to easily integrate new features in the pipeline is fundamental to comply with new data acquisition protocols, legacy systems or arising ODA use cases.
    \item \textbf{Flexibility}: an ODA pipeline must, by design, be adaptable to a wide variety of monitoring and ODA use cases, as well as to the respective operational needs.
\end{itemize}

Based on these requirements, \lrz has realized an end-to-end pipeline of tools and techniques, which we use for our deployments in Sections~\ref{section:sngdeployment} and~\ref{section:dpdeployment} to address the challenges posed by ODA. Our pipeline was developed from the ground up without any dependency on the specific infrastructure of our data center, with the aim of providing a generic solution suitable for most production use cases relying on time-series sensor data. It comprises the following three components:

\begin{figure}[t]
\centering
\captionsetup[subfigure]{}
\subfloat[Raw data.]{
\includegraphics[width=0.20\textwidth,trim={10 0 120 0}, clip=true]{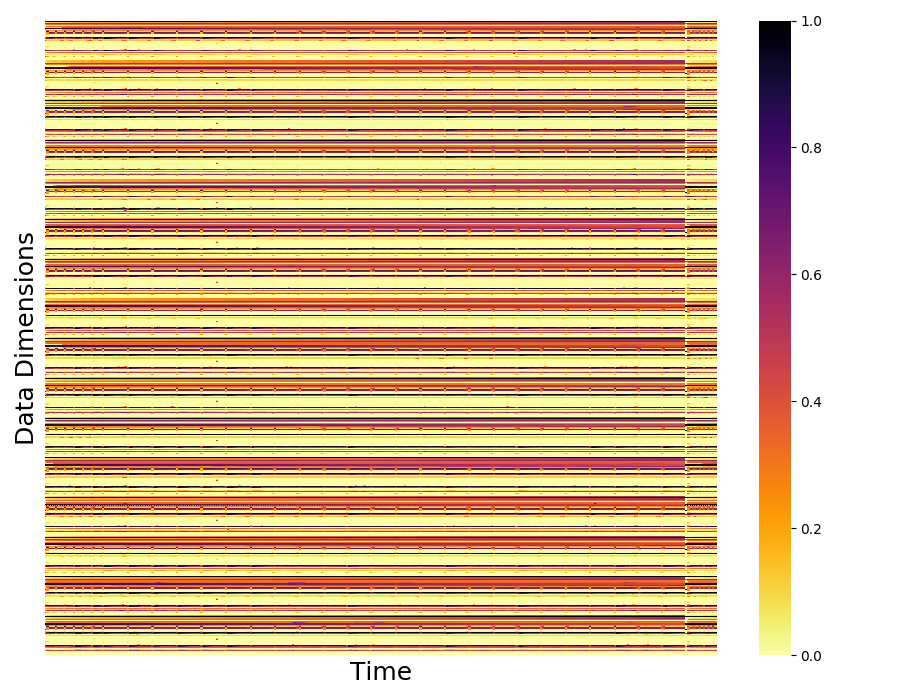}
  }
\subfloat[\csmethod signatures.]{
\includegraphics[width=0.20\textwidth,trim={10 0 120 0}, clip=true]{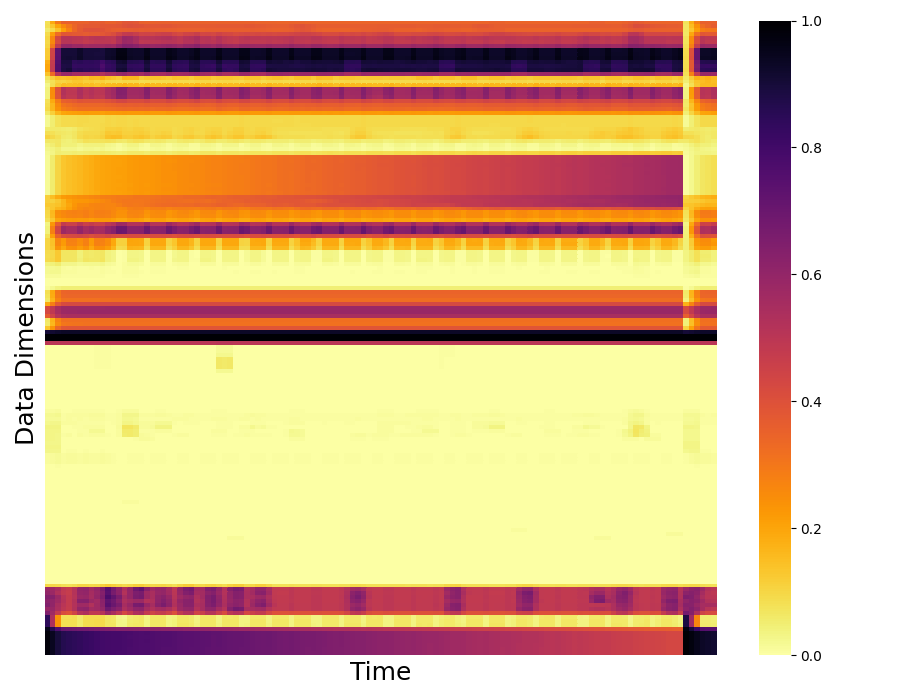}
}
\caption{Raw monitoring data ({\raise.17ex\hbox{$\scriptstyle\sim$}}800 sensors) and the resulting \csmethod signatures (160 blocks). Each column is a signature, and darker colors signify higher values.}
\label{method:correlationsorting}
\end{figure}

\begin{itemize}
    \item The \textbf{Data Center Data Base} (DCDB)~\cite{netti2019dcdb} is a \emph{scalable} and \emph{holistic} monitoring solution designed to handle large-scale data center installations: it is based on Pushers, which are \emph{modular} plugin-based daemons running in monitored entities and collecting data from various sources, such as CPU performance counters and infrastructure energy meters, in a uniform and \emph{comparable} format. Sensor data is transmitted using the \emph{Message Queuing Telemetry Transport} (MQTT) protocol~\cite{locke2010mq} to Collect Agents, which act as brokers forwarding all data to a persistent Storage Backend; in our case, this is a distributed Apache Cassandra noSQL data store~\cite{wang2012nosql}. 
    \item \textbf{\daf}~\cite{netti2019wintermute} is a generic ODA framework designed to be integrated into any monitoring solution, with operator plugins supplying arbitrary analysis or control capabilities in a \emph{flexible} manner. The currently available plugins allow for aggregation of data, computation of per-job metrics, output of sensors to arbitrary sinks, and machine learning tasks such as regression, classification and clustering. \daf is tightly integrated into the Pushers and Collect Agents of \dcdb, and operator plugins can be instantiated in both without code changes. Moreover, \daf supplies a series of \emph{abstraction} constructs (denoted as \emph{block system}) that allow to write compact configurations, even for complex ODA models deployed at a large scale.
    \item \textbf{Correlation-wise Smoothing} (CS)~\cite{netti2020correlation} is a monitoring data processing technique designed to \emph{compress} sensor data into compact signatures that are descriptive of a component's status. The signatures are built by arranging data dimensions on a bi-dimensional space, where the horizontal axis represents time and the vertical one the data dimensions, with the latter ordered according to their correlations. By applying smoothing on this space we are able to produce compact image-like signatures that are \emph{visualizable}, \emph{manipulatable}, \emph{portable} across systems and that exhibit good ODA \emph{performance}. This technique is implemented as a \daf operator plugin. Figure~\ref{method:correlationsorting} demonstrates its effect on monitoring data.
\end{itemize}

Having provided a broad picture of the state of ODA in HPC, in the next sections we demonstrate how we use the pipeline introduced above to address two complex ODAV and ODAC use cases on different production HPC systems, each with different requirements, scale and architecture. We will focus on the challenges, opportunities and lessons learned when transitioning ODA into production, with the goal of identifying common pitfalls and providing guidelines to ease such deployments in other centers.

%% file: sections/SNGDeployment.tex
\section{Job Data Visualization on \sng}
\label{section:sngdeployment}

The first ODA production use case is an ODAV deployment on the \sng HPC system at \lrz, tailored for online visualization of job-level performance metrics, which is in continuous production operation since September 2020.

\subsection{Overview}

With more than 25 PFlop/s of peak performance, \sng\footnote{\url{https://doku.lrz.de/display/PUBLIC/SuperMUC-NG}} is the flagship HPC system at \lrz and is among the 20 most powerful systems in the world as of November 2020\footnote{\url{https://top500.org}}. It comprises more than 6,000 compute nodes, distributed over 8 islands, and each equipped with two 24-core Intel Skylake CPUs, 96GB of RAM and an Intel Omni-Path fabric. The nodes are disk-less and employ the SLES 12 operating system, while a SLURM installation manages the cluster and GPFS provides the main file system.

Effective usage of such a large-scale system requires deep understanding of application behavior, which is in turn a challenging task due to the complexity, scale and duration of the job runs on a system of this kind. ODA simplifies this process considerably: the previous iterations of the \sng line employed the \persyst framework~\cite{Guillen2014}, which is an end-to-end solution to collect metrics of interest from compute nodes, store them in a database and finally visualize them on a per-job basis in a web frontend. \persyst relies on computing quantiles from the distributions of certain derived performance metrics, which can effectively highlight performance bottlenecks and inefficiencies in large-scale user jobs. However, previous iterations of \persyst performed monitoring at a coarse 10-minute granularity, which was to be improved on the new system. Due to the technical challenges associated with \sng's scale, we decided to decouple the \persyst web frontend, maintaining end user access, from the backend monitoring and aggregation of sensor data: this was entrusted to \dcdb and \daf in order to enable better scalability, maintainability as well as extensibility to other use cases within a single framework. These perform all of the necessary monitoring and data aggregation in a transparent way, feeding the processed data into a dedicated database and subsequently enabling visualization via the \persyst frontend.

\subsection{Monitoring Infrastructure}

\begin{figure}[t]
\centering
\includegraphics[width=0.41\textwidth,trim={0 0 0 0}, clip=true]{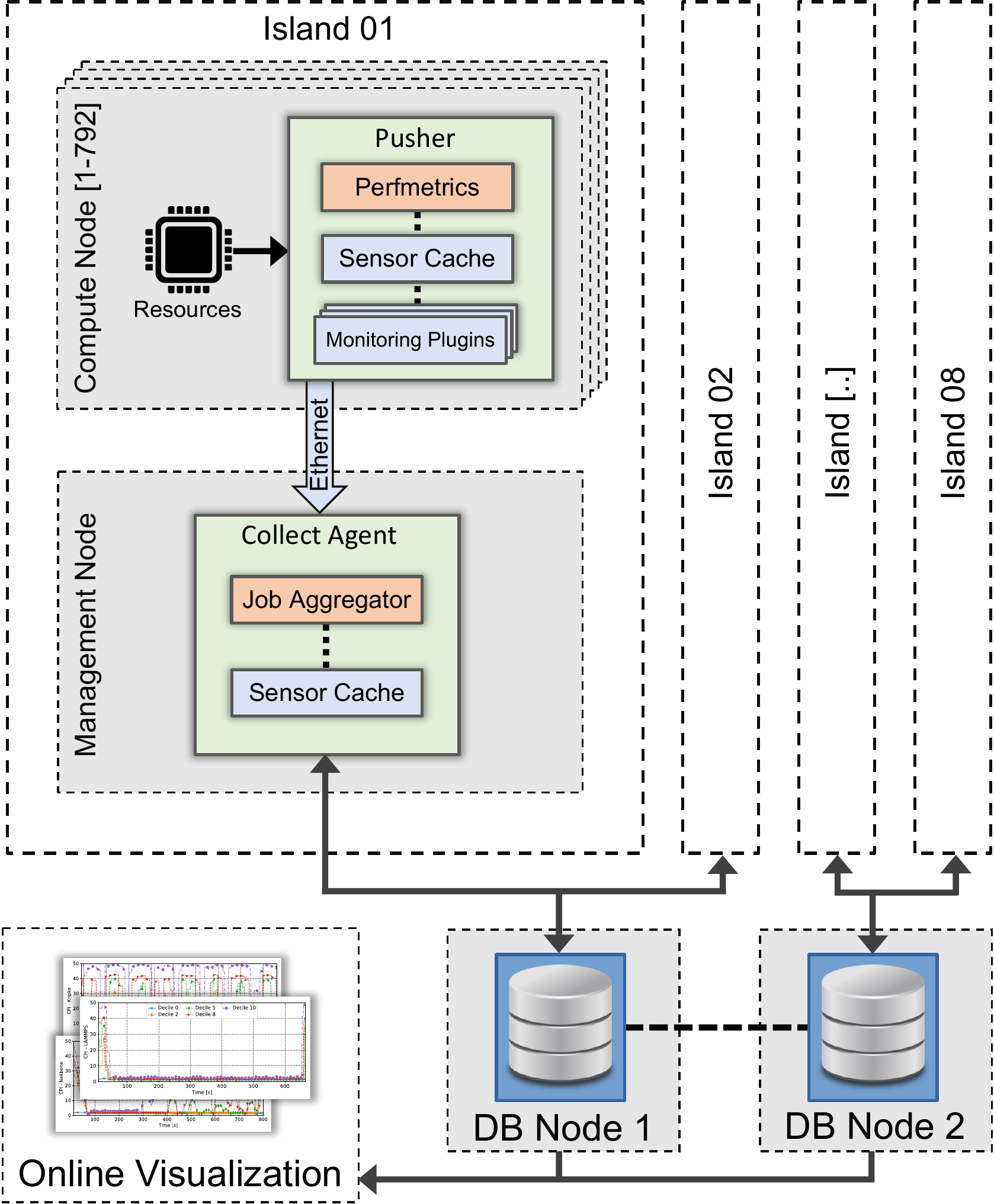}
\caption{A diagram representing the ODAV infrastructure deployed on \sng for job data visualization.}
\label{ng:deployment}
\end{figure}

The placement of \dcdb and \daf components on \sng is summarized in Figure~\ref{ng:deployment}. A \dcdb Pusher runs in each compute node, collecting monitoring data at 10s intervals using a series of plugins: \emph{Perfevent} and \emph{MSR} sample a variety of CPU performance counters (e.g., instructions or cache misses or cycles), while \emph{SysFS} collects CPU energy and temperature, as well as DRAM energy and several Omni-Path-related metrics from a virtual file system. Finally, \emph{ProcFS} samples memory and CPU utilization metrics from the \emph{stat} and \emph{meminfo} files, while \emph{GPFSMon} samples metrics associated with I/O activity on the GPFS file system. Each \sng island has a dedicated management server hosting a Collect Agent instance, to which Pushers in the same island send data via MQTT. This is done using a dedicated Ethernet interface for telemetry, so as not to interfere with the communication of user applications. The Collect Agents finally insert the data into a Storage Backend distributed on two nodes, each handling 4 islands, from which it can then be queried. We use a dual storage setup: a distributed Cassandra database stores most of the monitoring data, with a short time-to-live of 30 days; a \emph{MariaDB}\footnote{\url{https://mariadb.org}} instance, on the other hand, is used for accounting of user activity, as well as to store all \persyst data that is used for visualization, for the entire life of the system. All \dcdb binaries, libraries and configurations are stored on the GPFS file system and are accessible from all compute nodes, making it easy to deploy changes on short notice.

\subsection{ODAV Infrastructure}

We use \daf for our ODAV needs, employing two different operator plugins in a pipeline. This is once again shown in Figure~\ref{ng:deployment}: we use the \emph{Perfmetrics} plugin in the Pushers, and the \emph{Job Aggregator} plugin in Collect Agents - they operate respectively in-band and out-of-band. Both plugins operate at an interval of 2m, which was a significant improvement over the 10m of previous systems, while still resulting in very light storage requirements. In fact, all of the job data computed via \daf must be kept for the entire life time of the system and is never deleted. The information about jobs running on the system is exposed via a separate tool, which pulls data from SLURM and stores it in Cassandra.

The Perfmetrics plugin is configured to compute 27 derived node and CPU core-level metrics, picking the most recent value (related to the latest 10s) for each input sensor every 2m; the effectiveness of this type of statistical sampling was proven in the context of the original \persyst framework~\cite{Guillen2014}. The final set of metrics is able to characterize the performance of a user application, with indicators about CPU performance (e.g., \emph{Floating Point Operations per Second} or FLOPS) or memory activity (e.g., memory bandwidth), among others. The Job Aggregator plugin in the Collect Agents, on the other hand, is configured to aggregate each of the 27 derived metrics from the compute nodes associated with running user jobs, every 2m: for each metric, it computes the associated deciles, the average and a \emph{severity measure}, which is an efficiency indicator against a pre-defined threshold. The final per-job data is then pushed to the MariaDB instance for long-term storage, unlike all other data. The runtime overhead for monitoring~\cite{netti2019dcdb} and for the Perfmetrics plugin~\cite{netti2019wintermute} in compute nodes was proven to be well below 1\%. 

\subsection{Challenges and Solutions}

We now discuss the challenges associated with the large scale and integration requirements of this ODAV deployment, as well as the solutions we adopted.

\paragraph{In-memory Processing.}

Our configuration results in 14,5 millions of sensors system-wide, excluding the final per-job data. However, the raw sensors sampled in the Pushers (e.g., CPU counters), which amount to roughly 6,8 millions, are used solely to process derived metrics: for this reason, they are configured such that they are never sent to Collect Agents. Instead, they are kept in the Pushers' cache, which contains the most recent readings for each local sensor, so as to be leveraged by the Perfmetrics plugin - this is only possible due to the tight integration between \dcdb and \daf, which allows for direct in-memory processing of data. The derived Perfmetrics sensors are then sent out every 2m to Collect Agents, with 1,213 sensors per node. As the raw data is sampled every 10s, this leads to a much lighter burden on the transmission and storage systems, with 60,000 inserts per second into Cassandra, as opposed to the 700,000 in the naive configuration. Perfmetrics data has a time-to-live of 30 days, further reducing storage requirements down to 2.5TB for Cassandra on \sng. An interval of 2m, on the other hand, results in a monthly average of 10GB of per-job data added to MariaDB, which is sustainable for years to come.

\paragraph{Workload Distribution.}

Processing the FLOPS metric for a job running on 1,024 nodes requires fetching data for 49,152 distinct sensors, one for each CPU core associated with the job. This would result in several millions of database queries every 2 minutes to process all 27 metrics: as such, each Collect Agent only aggregates data for jobs which have the majority of their compute nodes in the same island as itself. This strategy is the most natural way to distribute work on \sng, as islands are equally sized. Moreover, this allows to optimize access to sensor data: since most compute nodes for a given job always belong to the same island as the Collect Agent, the respective Perfmetrics data will be readily available in the local cache, which contains recent sensor readings received via MQTT, reducing load to the Storage Backend. Sensors that are not available in the cache (e.g., for jobs spanning multiple islands) are queried from the Cassandra database transparently.

\paragraph{Integration with Legacy Tools.}

As the legacy web frontend for the \persyst framework supported only MariaDB as Storage Backend and no development was planned on this component, integration had to be done on the \dcdb side. The plugin-based and generic nature of \daf aided us in this purpose: in fact, we integrated all appropriate logic to manage MariaDB connections and perform insert operations within the Job Aggregator plugin itself - in general, a \emph{sink} \daf plugin is one that is able to write to alternative Storage Backends. This does not impair the plugin's ability to query sensor data from (as well as publish to) the default Cassandra database. This approach offered a clean and transparent upgrade path - no code changes were required to the \persyst, \dcdb or \daf cores, obtaining in turn an open-source plugin that might be useful for future use cases as well.

\subsection{Usage Example}

\begin{figure}[t]
\centering
\captionsetup[subfigure]{}
\subfloat[Job overview screen.]{
\includegraphics[width=0.465\textwidth,trim={0 19 150 0}, clip=true]{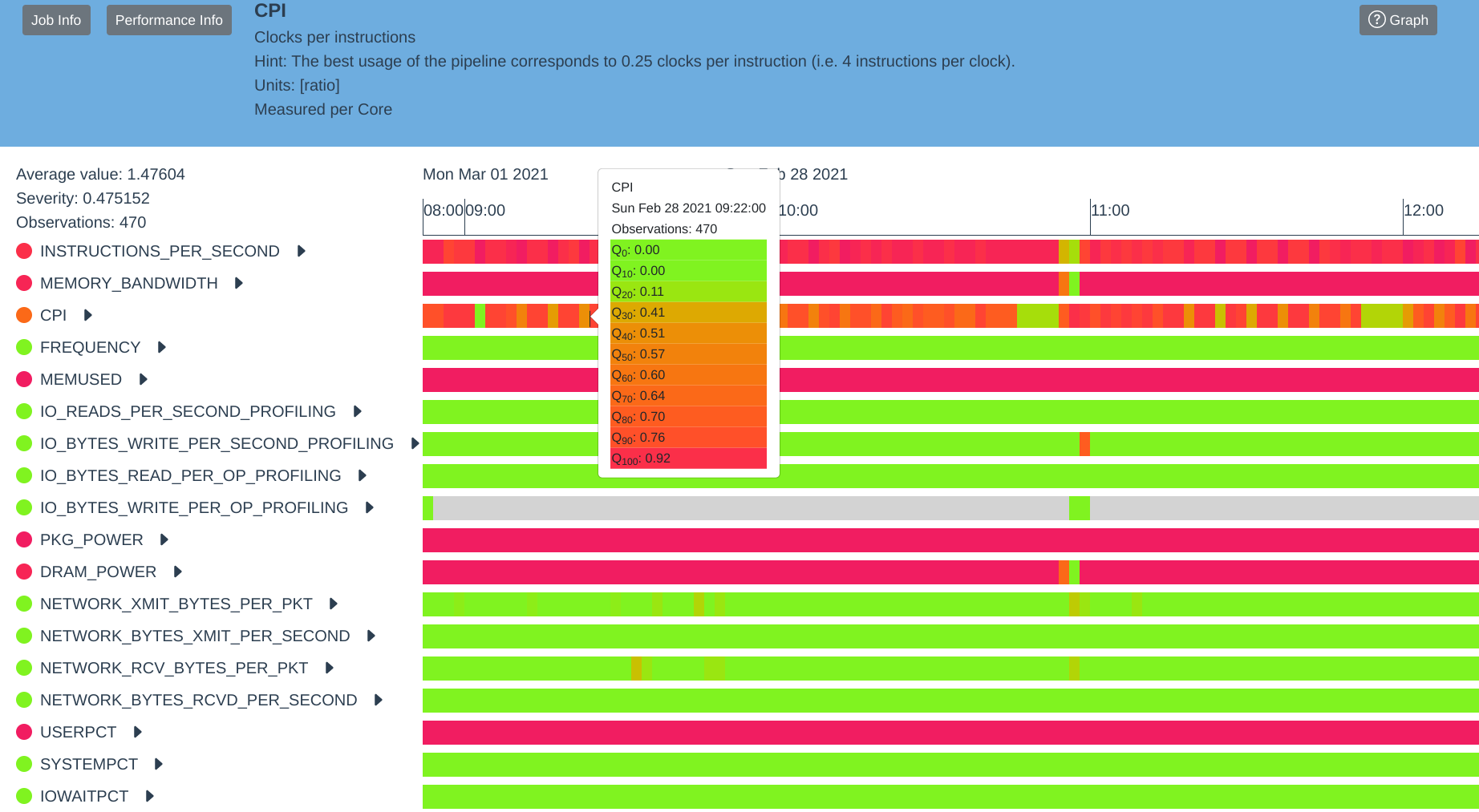}
  } \\
\subfloat[Detailed screen.]{
\includegraphics[width=0.465\textwidth,trim={0 30 100 0}, clip=true]{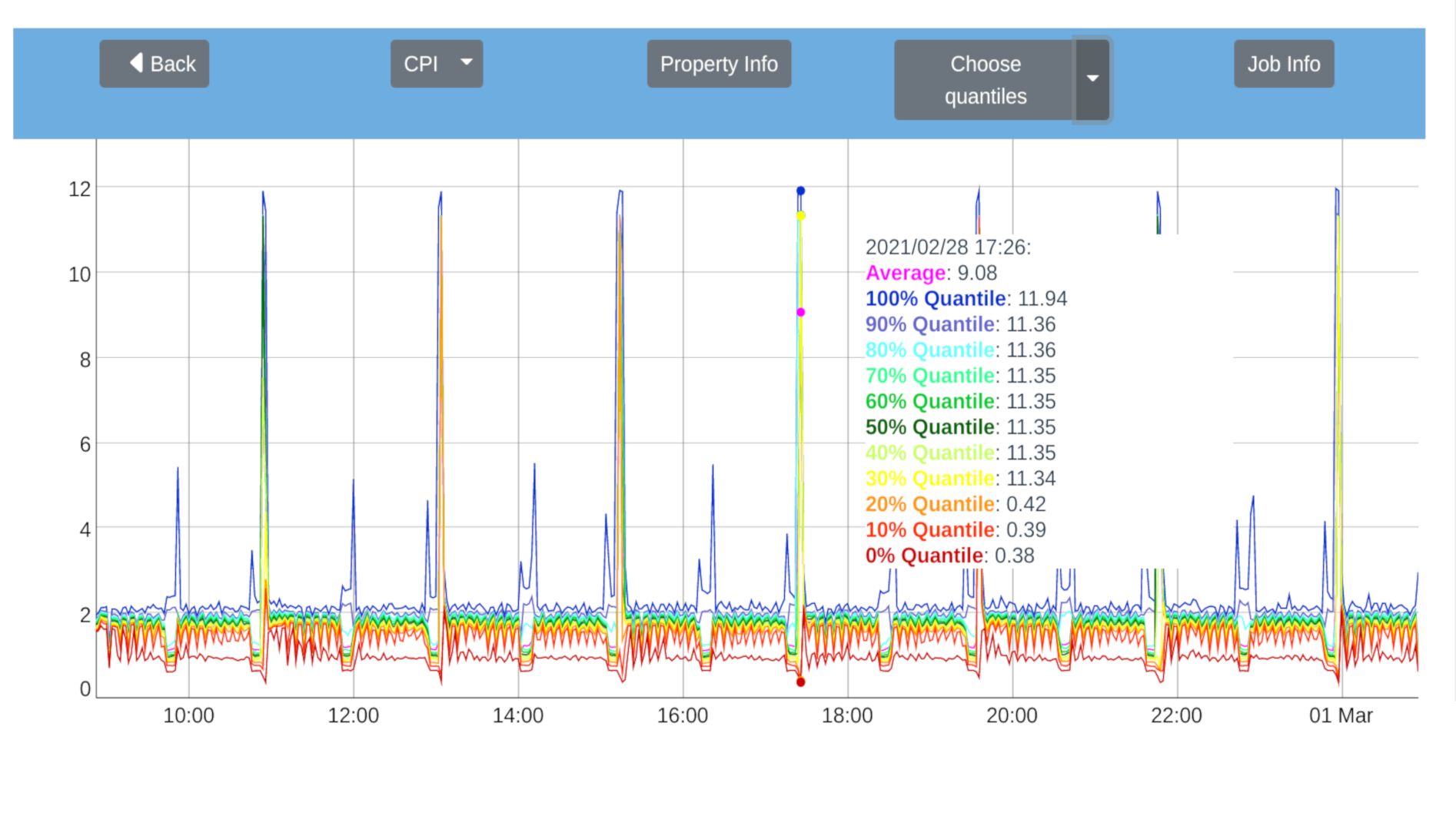}
}
\caption{Two screens from the \persyst web frontend.}
\label{ng:persystscreen}
\end{figure}

Here we provide a brief overview of how the per-job aggregated data stored in the MariaDB database through our ODA pipeline can be visualized by HPC users on the \persyst web frontend. An extensive analysis of the performance metrics exposed to users and their meaning was the object of previous work~\cite{Guillen2014,brayford2020analyzing}. Figure~\ref{ng:persystscreen} shows an overview of the \persyst interface: users are required to log into the frontend using the same credentials as on the \sng system, leading to an initial web page containing a list of jobs submitted by the authenticated user whose data can be visualized - jobs do not necessarily have to be finished for their data to be visualized, and they can be analyzed as they run, online. 

Selecting a job leads to an overview page, as shown in Figure~\ref{ng:persystscreen}a: here, a global view of the job's performance is presented, showing the medians of all available performance metrics over time in a heatmap-like style. Each pixel of the heatmap shows the median of a certain performance metric over a specific 2-minute time window, as aggregated by the \daf Job Aggregator plugin from all compute nodes associated with the job. Placing the cursor over any of the data points highlights the full distribution (shown in terms of deciles) of the selected performance metric in the given time range, allowing to spot performance anomalies and imbalances with ease. Clicking on any of the metric names, finally, leads to a new page (shown in Figure~\ref{ng:persystscreen}b) which allows users to visualize the deciles and average of the selected metric over time for a more detailed analysis. In this type of visualization, multiple metrics at a time can be shown for comparison purposes. In this example, the deciles and average of the \emph{Cycles Per Instruction} (CPI) metric are visualized over time, highlighting a clear periodic behavior that is likely associated with the job's underlying HPC application cycling between compute and communication-intensive phases. On top of using the web frontend, users can also query all available data via the \dcdb command-line tools.

%% file: sections/DPDeployment.tex
\section{Predictive Cooling Control on the \dpp System}
\label{section:dpdeployment}

The second ODA production use case is an ODAC deployment on the \dpp system at \jul, used to perform predictive optimization of the system's warm-water cooling infrastructure. It is in active use since October 2020.

\subsection{Overview}

The \dpp HPC system\footnote{\url{https://fz-juelich.de/ias/jsc/EN/Expertise/Supercomputers/DEEP-EST/_node.html}}, hosted by \jul, is composed of 3 heterogeneous \emph{modules}, each with a different architecture and functionality. The \emph{Cluster Module} (CM) comprises 50 nodes within a single rack, each equipped with two 12-core Intel Skylake CPUs, 192GB of RAM and a Mellanox Infiniband interconnect. The \emph{Extreme-Scale Booster} (ESB), on the other hand, is composed of 75 nodes in 3 separate racks - these employ an 8-core Intel Cascade Lake CPU supported by a Nvidia V100 GPU, coupled with 48GB of RAM and an Extoll interconnect. Finally, the \emph{Data Analytics Module} (DAM) comprises 16 compute nodes in a single rack, which employ two 24-core Intel Cascade Lake CPUs, a Nvidia V100 GPU and an Intel Stratix 10 FPGA, plus 384GB of RAM and an Extoll interconnect. The \CM and \ESB are warm-water-cooled, while the \DAM is air-cooled; a set of gateway nodes manages inter-module communication. Each compute node is fitted with an SSD and employs the CentOS 7 operating system, while BeeGFS and GPFS instances support the file system. As the \dpp modules are optimized for different workloads, the entire system is managed as a single SLURM cluster and dedicated options are supplied for jobs to run as \emph{workflows} over multiple modules~\cite{erlingsson2019scalable}.

We were tasked with supplying monitoring for the system as well as insightful data analytics using the \dcdb and \daf frameworks in the context of the wider \dpp research project. \daf implements an ODAC feedback loop that leverages predicted CPU and GPU temperatures for tuning the water temperatures of the \CM and \ESB racks, which employ direct-liquid warm-water cooling. The pipeline aims to keep inlet temperatures as high as possible at all times, while ensuring operational safety; this leads to an improvement of the cooling system's efficiency and allows for re-use of waste heat, for example via \emph{adsorption chilling} or through heating of office spaces~\cite{conficoni2015energy, wilde2017coolmuc}. The modular nature of the system and the different thermal profiles of its components call for a granular, rack-level control strategy, whose design takes inspiration from previous work~\cite{jiang2019fine}. On top of this ODAC pipeline, \daf also supplies per-job aggregation of data as done in Section~\ref{section:sngdeployment}, showing the benefit of using a single framework capable of supporting multiple concurrent use cases.

\subsection{Monitoring Infrastructure}

\begin{figure}[t]
\centering
\includegraphics[width=0.45\textwidth,trim={0 0 0 0}, clip=true]{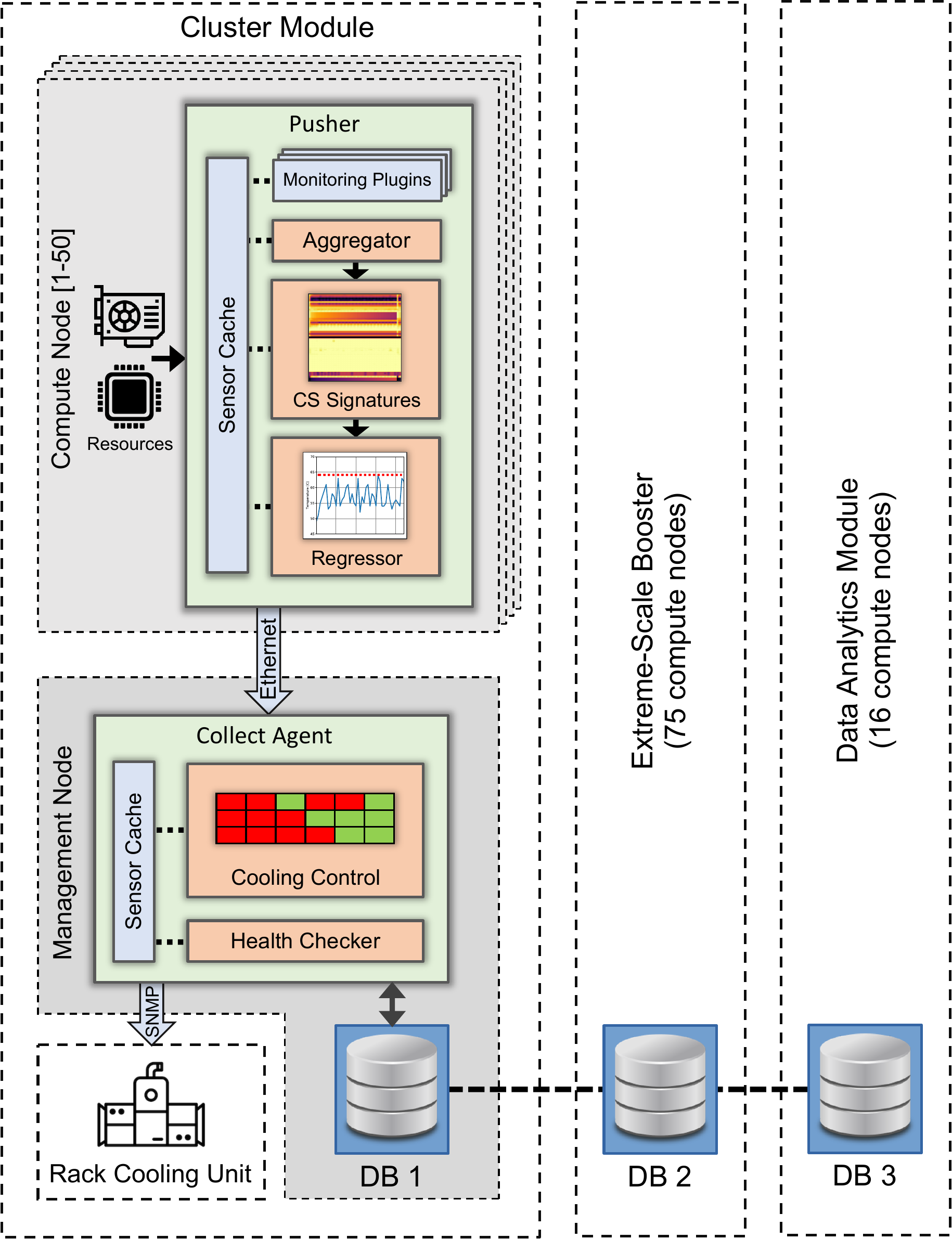}
\caption{A diagram representing the ODAC pipeline implemented on the \dpp system for cooling control.}
\label{deepest:coolingcontrol}
\end{figure}

In Figure~\ref{deepest:coolingcontrol} we show the placement of \dcdb and \daf software components over the \dpp system. Once again, one Pusher daemon runs in each compute node: due to the heterogeneity of the \CM, \ESB and \DAM, the respective Pushers have different configurations. Overall, we use the Perfevent plugin for sampling of CPU performance counters, ProcFS for resource utilization metrics from the stat, meminfo and \emph{vmstat} files, and SysFS for a variety of metrics. These include node and CPU energy and temperature, as well as DRAM and GPU energy (if available) and performance counters from the Infiniband or Extoll interconnect. In addition to these plugins, we use the \emph{NVML} plugin on \ESB and \DAM nodes to sample a variety of GPU-related metrics, such as utilization and clock. All plugins use a sampling interval of 10s and, due to the modest size of this system, the time-to-live of the sensors - roughly 70,000 in total - is set to one year.

Pushers transmit their data via MQTT to Collect Agents running on dedicated management nodes, one for each module. These same nodes also host a distributed Apache Cassandra instance, with each node storing the monitoring data of the corresponding module. As done previously, communication occurs via dedicated Ethernet interfaces to minimize overhead on user applications. An additional Pusher running on one of the management nodes collects data from the warm-water cooling and energy infrastructure via the \emph{SNMP} and \emph{REST} plugins. Each \CM and \ESB rack has its own \emph{Rack Cooling Unit} (RCU) managing the flow of water from and to the building infrastructure, that is exposed via an SNMP interface for querying operational parameters and enacting control. All \dcdb components are installed as RPM packages.

\subsection{ODAC Infrastructure}

The pipeline of \daf plugins used on the \dpp system is once again shown in Figure~\ref{deepest:coolingcontrol}. In all of the \CM, \ESB and \DAM Pushers there is an \emph{Aggregator} operator plugin running, which creates node and socket-level aggregates of per-CPU performance counters at 10s intervals. Then, \CM and \ESB Pushers implement a pipeline to predict CPU and GPU temperatures, used to steer control decisions for the associated RCU. This is supported by two operator plugins: first, a \emph{\csmethod} operator plugin provides an online implementation of the homonymous technique~\cite{netti2020correlation}, which converts the data associated with a set of input sensors (both raw and produced by the Aggregator plugin) into a compact signature made of 20 \emph{blocks} (i.e., complex coefficients). The resulting 40 sensors are then leveraged by a \emph{Regressor} plugin, which implements random forest-based regression using the \emph{OpenCV} library\footnote{\url{https://opencv.org}} - here, an ML model is configured to use as feature vectors the \csmethod signatures and produce as output a prediction of the \emph{maximum} expected CPU or GPU temperature in the near future. Both plugins operate at 1m intervals and we use three distinct models: one for \CM CPUs, one for \ESB CPUs and another for \ESB GPUs. This part of the pipeline is in-band and operates within local memory.

The pipeline continues in the \CM and \ESB Collect Agents with the out-of-band \emph{Cooling Control} operator plugin: it is tasked with determining new settings periodically for the secondary (i.e., of the rack-internal loop) inlet water temperature of each RCU, or its \emph{set temperature}, defined as \(T_{rcu}\). In practice, changing this parameter eventually causes the controller within the RCU to open or close an internal valve, affecting the flow of cold water coming from the building infrastructure, and hence the temperature of the water reaching the compute nodes. It would be preferable to set the valve's position directly instead of having the RCU's controller react to the change of the set temperature with some delay. However, as this setting is not exposed via the SNMP interface, the temperature prediction compensates for the latency of the controller at cooling demand changes. The plugin works at 1m intervals, providing a good compromise between overhead and control granularity, as in the following: for each RCU, it queries the predicted temperature of each CPU or GPU \(i\) associated with it, which can be simply fetched from the Collect Agent's sensor cache. If the prediction exceeds a threshold $T_{hot}^i$, the component is counted as \emph{hot}, and hence in need of cooling. The algorithm then computes the fraction $P_{hot}$ of hot components over the total, and updates \(T_{rcu}\) as follows:

\begin{equation}
    T_{rcu} = T_{rcu} + (T_{max} - T_{min})\cdot(P_{th} - P_{hot})
\label{coolingcontroleq}
\end{equation}

In Equation~\ref{coolingcontroleq}, $T_{min}$ and $T_{max}$ are respectively the minimum and maximum possible \(T_{rcu}\) settings, while $P_{th}$ is the configurable fraction of components that are allowed to be hot at a given time: if $P_{hot}$ is higher than this, \(T_{rcu}\) will be lowered, and conversely it will be increased. The new $T_{rcu}$ setting is finally applied to the RCU via the SNMP protocol. The plugin also supports the definition of critical thresholds \(T_{crit}^i\) which, if exceeded, lead to an immediate decrease of $T_{rcu}$ to $T_{min}$. The median runtime overhead of Pushers on single-node runs of the \emph{High-Performance Linpack} (HPL) benchmark~\cite{dongarra2003linpack} was found to be below 1.1\% on all \dpp modules, with the Regressor plugin's impact being negligible~\cite{netti2019wintermute}. Finally, all of the Collect Agents (\DAM included) leverage the Job Aggregator plugin for the computation of per-job metrics, plus a \emph{Smoothing} plugin that computes aggregates at 5m and 60m intervals for most sensors - these are not shown in Figure~\ref{deepest:coolingcontrol} for space reasons. The aggregates produced by these two plugins are not subject to a time-to-live of one year and are kept for the lifetime of the system. This leads to 3.5TB of Cassandra storage for ordinary \dpp sensor data, plus roughly 3GB per month for the aggregates.

\subsection{Challenges and Solutions}

We now highlight the challenges of this ODAC use case, focusing on the complexity of using a machine learning-based control pipeline in an heterogeneous HPC system.

\paragraph{From In-band Data to Out-of-band Control.}

Orchestrating a system's operation with data and knobs at multiple levels is one of the main challenges of ODAC. \daf's holistic nature and generic interfaces allow us to overcome this problem: it enables us to connect compute node-level data processed in-band (through the Regressor plugin) to subsequent out-of-band processing in the Collect Agents, which ultimately translates into new settings for SNMP system knobs and into a complete feedback loop. Further, to simplify the deployment of ML models, we leverage the plugins' ability to load model data from files: in particular, we train the \csmethod model (describing the sensors' permutation vector, as well as their lower and upper bounds) and the Regressor model (describing an OpenCV random forest) offline using archived data, which was processed via a Python framework; the final models are saved as files, which can be loaded by the C++ \daf implementations. If further adjustments to the models are required, online training sessions can be triggered via the \daf RESTful API~\cite{netti2019wintermute}.

\begin{figure}[t!]
\centering
\caption{An excerpt of the Cooling Control plugin's configuration for the \dpp \CM rack, using the block system.}
\label{fig:coolconf}
\begin{lstlisting}[linewidth=0.47\textwidth]
controller c1 {
  default def1
	
  input {
    sensor "<topdown 3, filter cm/s../socket>temp-p" {
      hotThreshold  73000
      critThreshold 93000
    }
  }
}
\end{lstlisting}
\end{figure}

\paragraph{System Heterogeneity.}

Configuring ODAC models on a modular system such as \dpp is not sustainable at scale: in our case, each RCU requires predicted temperatures from a specific set of components (i.e., CPUs or GPUs) within certain compute nodes. However, we are able to leverage the abstraction capabilities of \daf's block system, which arranges the space of available sensors in a tree-like structure based on their MQTT topics~\cite{netti2019wintermute}. Specification of the Cooling Control plugin's input sensors is therefore done using compact template-like expressions, which allow to pick specific sets of nodes and components with ease. This is shown in Figure~\ref{fig:coolconf}, which contains an excerpt of the Cooling Control plugin's configuration for the \CM module. A single template sensor is instantiated, specifying the \(T_{hot}^i\) and \(T_{crit}^i\) parameters for the \CM's CPU type. The \emph{sensor expression} at Line 5 specifies which sensors in the hierarchy should match the template: in this case, these are the \emph{temp-p} sensors associated with CPUs of \CM nodes.

\paragraph{Handling of Notifications.}

Automatic notification of anomalous events to system administrators is paramount for sustainable operation. In the context of \dpp, we use a \daf \emph{Health Checker} operator plugin in the \CM and \ESB Collect Agents, which verifies that the real temperature of each CPU or GPU never exceeds its \(T_{crit}^i\) value, at 1m intervals: if this happens, administrators are notified via e-mails. In general, the plugin supports the definition of several alert conditions (e.g., sensor data is above or below a certain threshold), as well as of arbitrary shell commands to handle and propagate notifications (e.g., to log streams). \daf's block system, as exemplified in Figure~\ref{fig:coolconf}, further allows to define alert conditions on thousands of sensors with very compact configurations. As \dcdb is fully integrated with the Grafana tool, the alerting capabilities of the latter can also be used with ease, while other tools can leverage \dcdb data via its command-line and RESTful interfaces for alerting purposes.

\subsection{Operational Results}

Here we discuss the effectiveness of CPU and GPU temperature prediction, as well as the impact of the Cooling Control plugin on infrastructure components of the \dpp system. In order to obtain reliable training data for the prediction models, we performed tailored 1-day experiments: for each of them, we leveraged real user jobs running on the \dpp modules, including a variety of CPU and GPU workloads, plus a series of test applications. These are \emph{Kripke}, \emph{AMG}, \emph{LAMMPS}, \emph{Quicksilver}, \emph{Nekbone} and \emph{PENNANT} from the \emph{CORAL-2} suite\footnote{\url{https://asc.llnl.gov/coral-2-benchmarks}}, plus the HPL benchmark. On the \ESB we also use the GPU-accelerated versions of LAMMPS, Quicksilver and HPL. Each application is configured to run under 3 possible input sizes on 16 compute nodes, with different MPI and OpenMP configurations. Throughout the duration of the experiments, we also apply several \(T_{rcu}\) settings, in the $[35C, 45C]$ range, for the \CM and \ESB RCUs. We then fetch monitoring data from compute nodes in the selected module using our production \dcdb installation, and use this to build the models' training sets. We adopted a similar approach to evaluate our cooling control pipeline, running several 1-day experiments each using certain $T_{hot}^i$ settings.

\subsubsection{Temperature Prediction Results}
\label{section:tempprediction}

\begin{figure}[b]
\centering
\includegraphics[width=0.475\textwidth,trim={10 0 10 0}, clip=true]{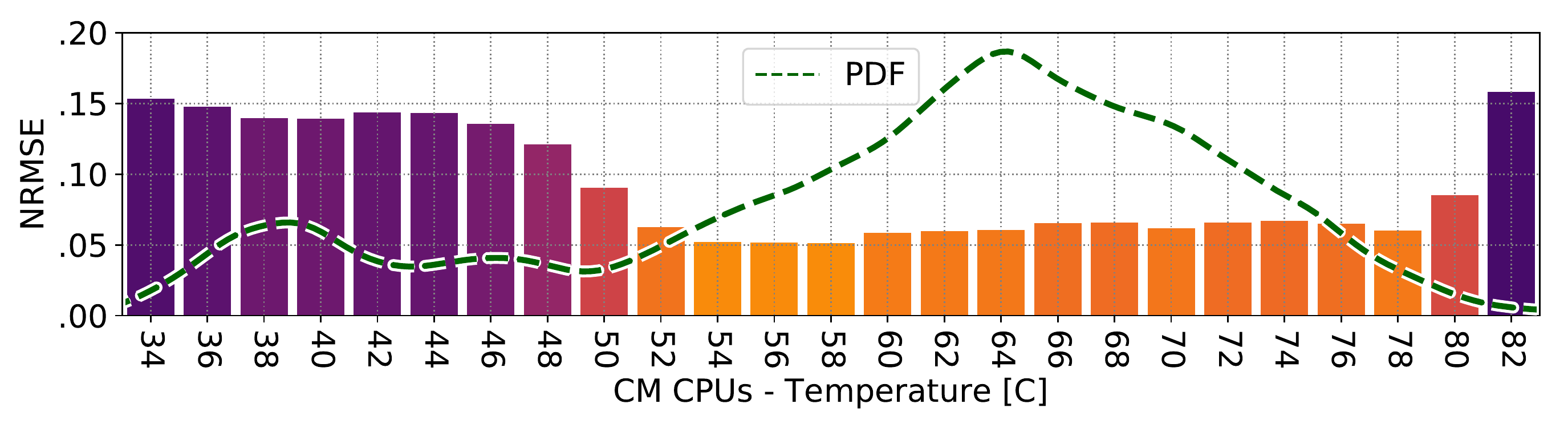} \\
\includegraphics[width=0.475\textwidth,trim={10 0 10 0}, clip=true]{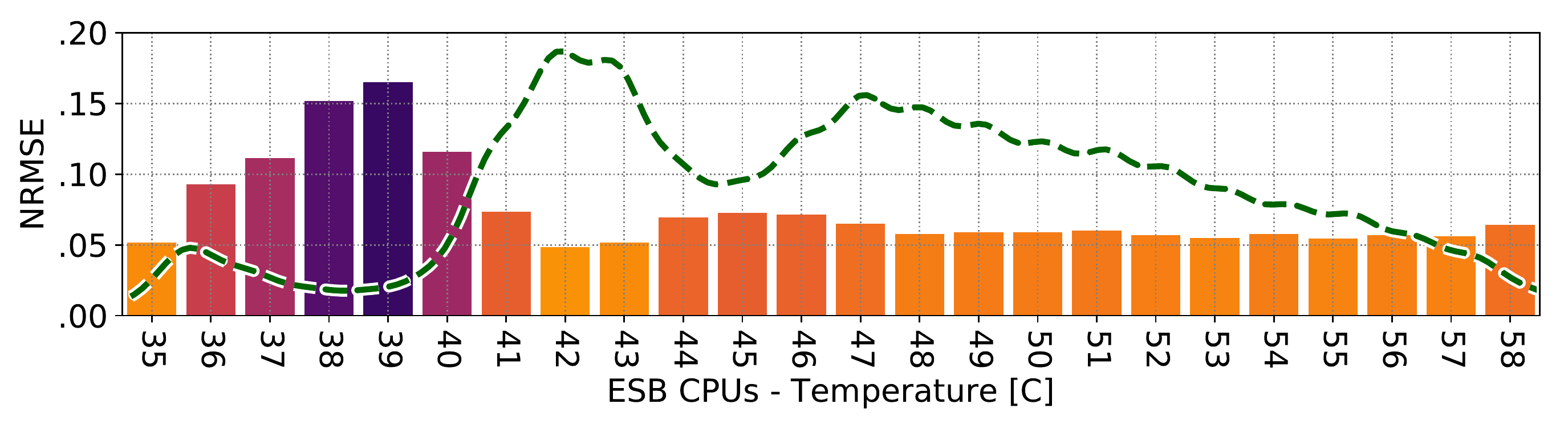} \\
\includegraphics[width=0.475\textwidth,trim={10 0 10 0}, clip=true]{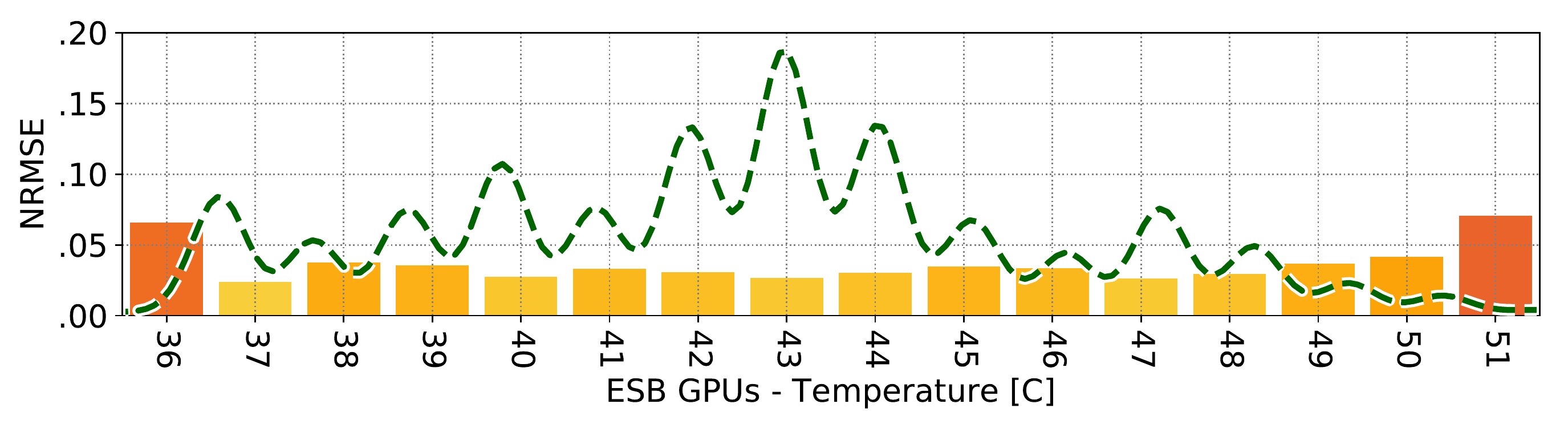}
 \caption{Prediction NRMSE at different temperature bands for the \CM CPUs, \ESB CPUs and \ESB GPUs models, together with the fitted PDFs of the original data.}
\label{deepest:coolingerr}
\end{figure}

We train each of the three temperature prediction models (one per component type) with data coming from a representative subset of 16 compute nodes, obtained by running two experiments on the \CM and \ESB as described above. For each model and component type we select a set of sensors (35 for \CM CPUs, 32 for \ESB CPUs and 27 for \ESB GPUs) and process this data using the \csmethod method: this gives us roughly 8,000 signatures per compute node (or twice for the dual-socket \CM nodes), which we merge into a single dataset and which we use as feature vectors. We then evaluated the effectiveness of prediction via 5-fold cross-validation, using the random forest implementation of the Python \emph{scikit-learn} library, with 50 estimators. After an initial exploration, we found the best configuration for the three models to use 20 \csmethod blocks (i.e., 40 coefficients) computed from the last 6 samples of sensor data (i.e., 1m), and predicting the maximum temperature in the upcoming 6 samples. We found one minute of prediction to be enough for the cooling system to fully react to changes in \(T_{rcu}\), while not resulting in over-fitting of the model. A counter-intuitive result is that training a global model with data coming from the selected 16 compute nodes, on top of being easier to maintain, also leads to substantially better performance in comparison to using per-node models: this is likely due to manufacturing variation among components~\cite{inadomi2015analyzing} introducing a healthy amount of noise during training. It should be noted that we did not observe any effects due to changes in \emph{leakage currents} on \CM and \ESB CPUs at different \(T_{rcu}\) settings, with identical energy consumption for the same workload at different temperatures; \ESB GPUs, on the other hand, show up to 5\% variation when using 35C or 45C as \(T_{rcu}\), which should be taken into account if targeting a large-scale GPU-based system with our control strategy. The dataset we use is freely available\footnote{\url{https://zenodo.org/record/4671477}} to ensure reproducibility.

In Figure~\ref{deepest:coolingerr} we show the bar plot of the \emph{Normalized Root Mean Square Error} (NRMSE) associated with each temperature band for the three models. We also show the fitted \emph{Probability Density Function} (PDF) of the original temperature data. The first clear observation is that \CM CPUs exhibit a much wider range of temperatures compared to \ESB CPUs and GPUs, which is due to their \emph{Thermal Design Power} (TDP) of 165W, as opposed to the 85W of \ESB CPUs; \ESB GPUs, on the other hand, exhibit the best thermal results despite their TDP of 250W. This has a direct impact on model performance, with the \CM CPU model exhibiting a global NRMSE of 0.077, while the \ESB CPU and GPU models sit respectively at 0.071 and 0.029. However, for all models the NRMSE is close to 0.05 for the most common temperature states, proving their effectiveness. Further, the models are naturally biased since they are trained to predict the maximum temperature rather than the average, leading to a natural tendency to over-estimation: this is a deliberate choice in order to make the control pipeline more robust against over-heating. Finally, using the \csmethod method gives us compact models that have a negligible impact on overhead and that are resistant against changes in the available set of sensors over time: after proving its validity through tailored experiments in previous work, this is the first real-life use case for this technique~\cite{netti2020correlation}.

\subsubsection{Impact on Infrastructure}

\begin{figure}[t]
\centering
\captionsetup[subfigure]{}
\subfloat[\CM.]{
\includegraphics[width=0.23\textwidth,trim={0 0 10 0}, clip=true]{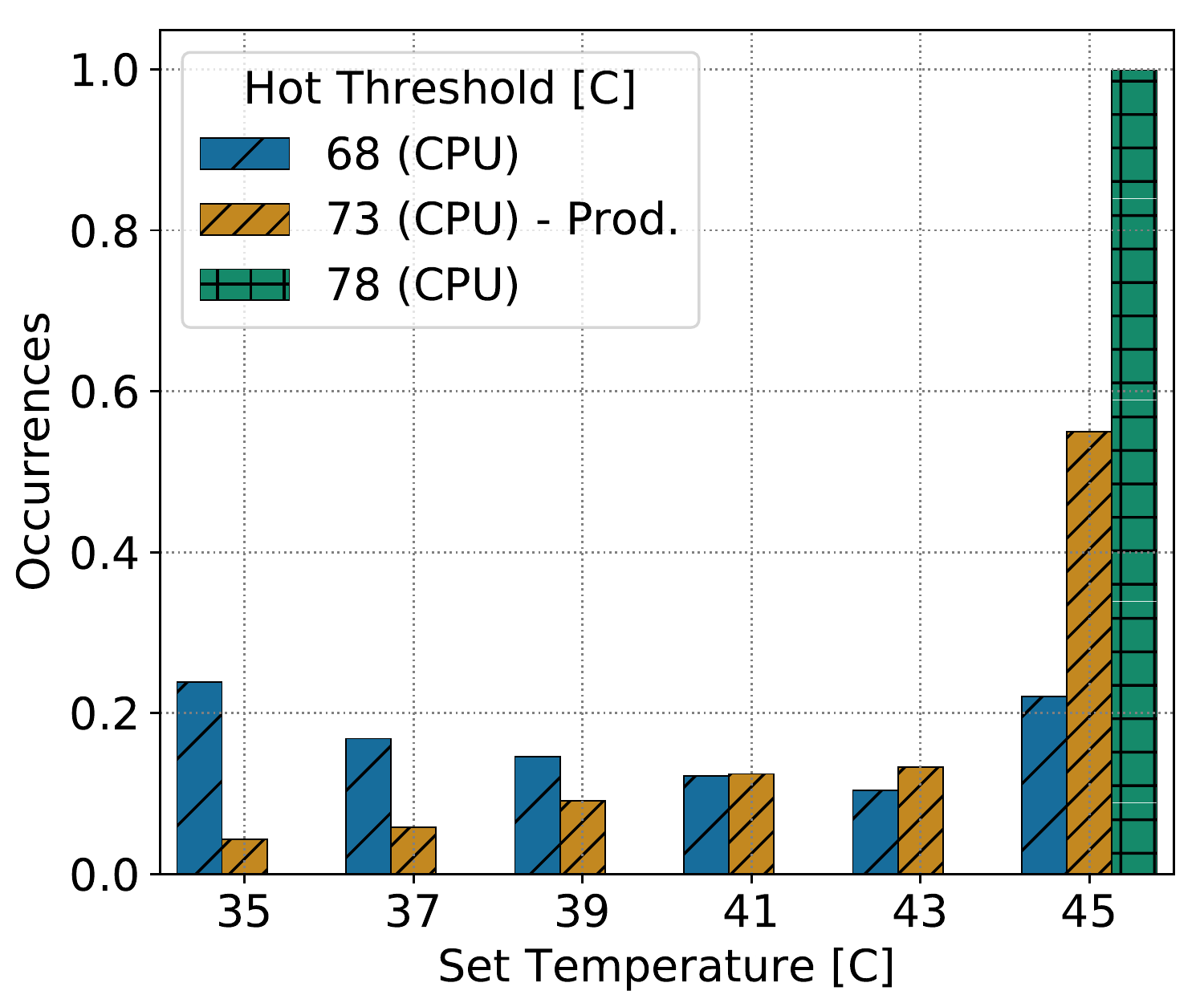}
  }
\subfloat[\ESB.]{
\includegraphics[width=0.23\textwidth,trim={0 0 10 0}, clip=true]{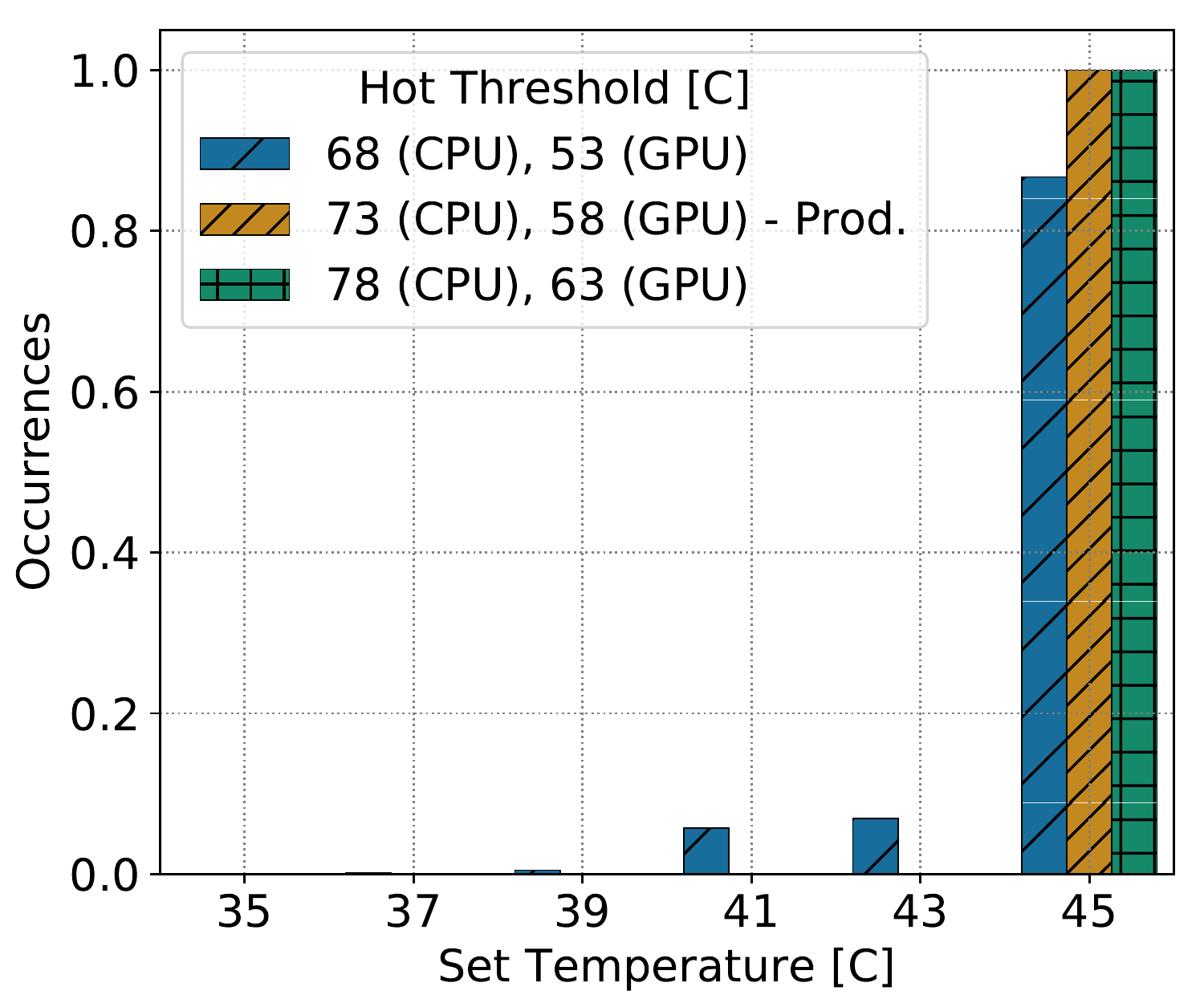}
}
\caption{Histograms of \(T_{rcu}\) values for the \CM and \ESB RCUs when using our control pipeline with different \(T_{hot}^i\) values.}
\label{deepest:coolinghist}
\end{figure}

We now discuss several 1-day experiments we performed on the \CM and \ESB with our ODAC pipeline, to highlight its behavior during production operation - we consider only one of the three identical \ESB racks. The Cooling Control plugin is configured as follows: we use a $P_{th}$ of 0.2, while $T_{min}$ and $T_{max}$ are set to 35C and 45C respectively. We experiment with different $T_{hot}^i$ values: 68C, 73C and 78C for \CM and \ESB CPUs, and 53C, 58C and 63C for \ESB GPUs. $T_{crit}^i$ is instead set to the thermal throttling temperature, which is 98C for \CM and \ESB CPUs, and 83C for \ESB GPUs. Thanks to the ample headroom between $T_{hot}^i$ and $T_{crit}^i$, our results can be generalized to more extreme configurations, by increasing both $T_{hot}^i$ and $T_{max}$. In Figure~\ref{deepest:coolinghist} we show the histogram of \(T_{rcu}\) values chosen by the Cooling Control plugin over the three \CM and \ESB experiments: different $T_{hot}^i$ settings result in different behaviors, with a transition from a uniform distribution at 68C, to an almost fixed 45C setting at 78C on the \CM. As we observed in Section~\ref{section:tempprediction}, the \ESB shows a less dynamic behavior and mostly stays at 45C under all settings, suggesting that operation at a higher $T_{max}$ would be possible. For long-term production operation, we use a $T_{hot}^i$ setting of 73C on \CM and \ESB CPUs, coupled with 58C for \ESB GPUs.

\begin{figure}[b]
\centering
\captionsetup[subfigure]{}
\includegraphics[width=0.475\textwidth,trim={10 0 10 0}, clip=true]{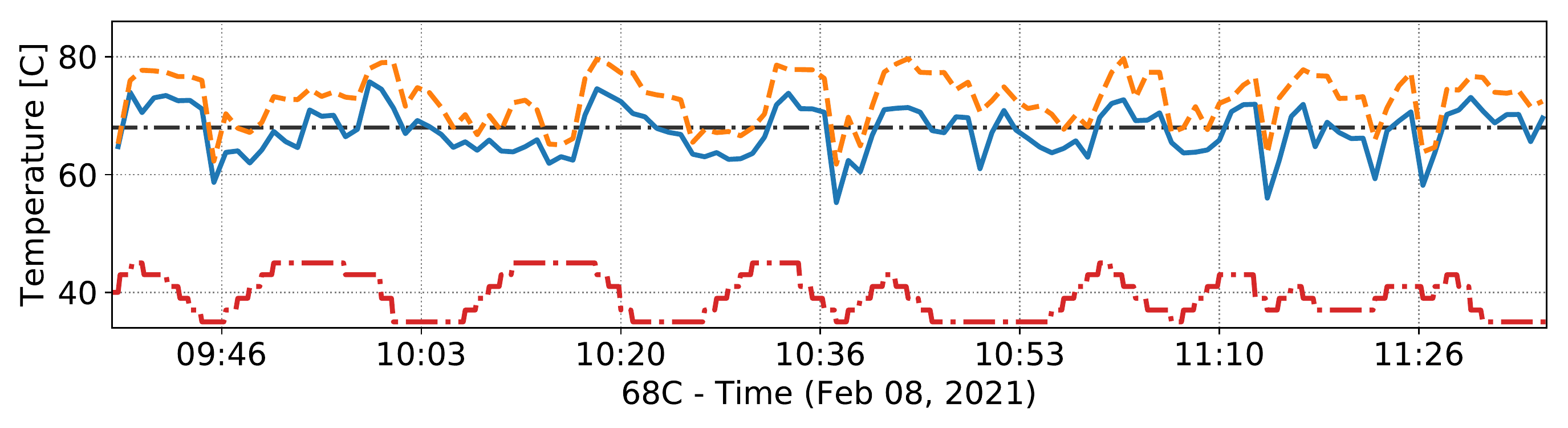} \\
\includegraphics[width=0.475\textwidth,trim={10 0 10 0}, clip=true]{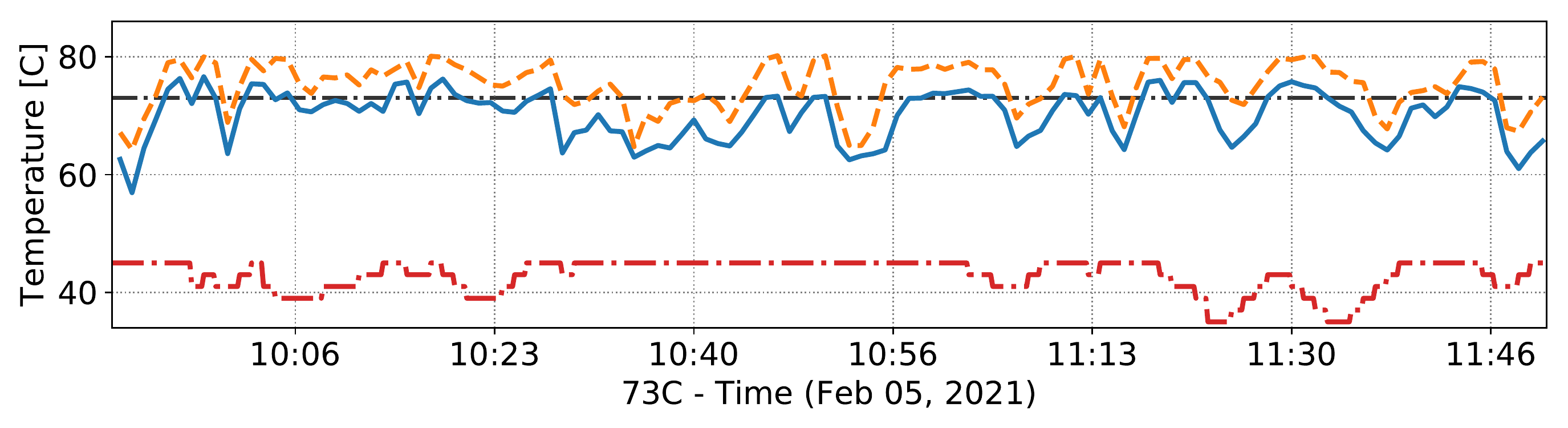} \\
\includegraphics[width=0.475\textwidth,trim={10 0 10 0}, clip=true]{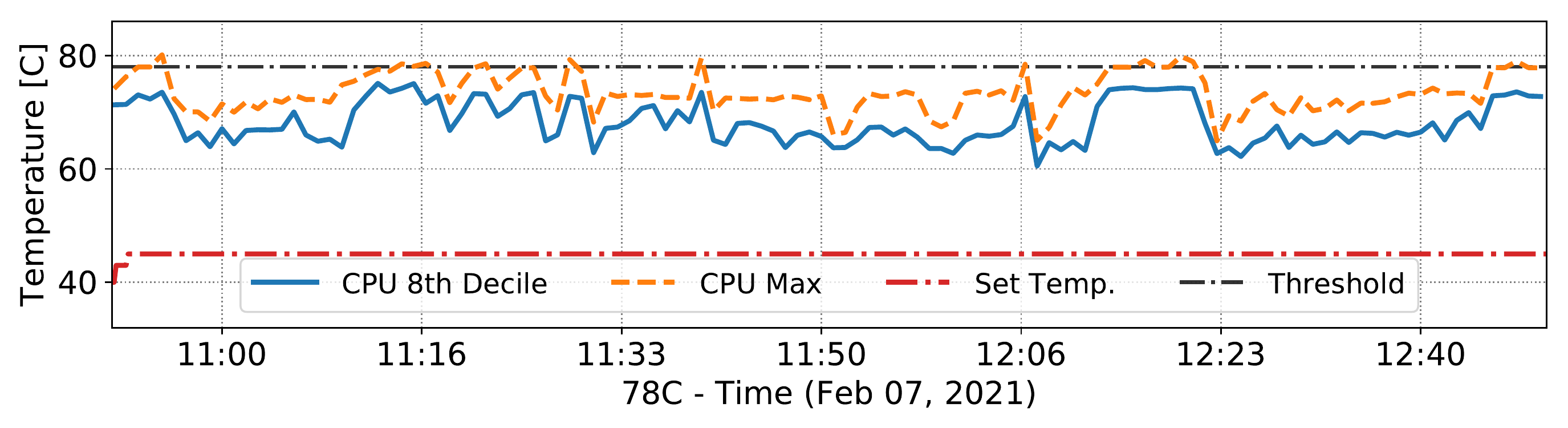}
\caption{Interaction between the \CM's \(T_{rcu}\) and CPU temperatures when enacting control with different \(T_{hot}^i\) values.}
\label{deepest:coolingtime}
\end{figure}

We now focus on the \CM in order to highlight the interactions between the control logic and the \dpp components. Figure~\ref{deepest:coolingtime} shows the \CM's \(T_{rcu}\), coupled with the 8th decile and maximum of predicted CPU temperatures from all nodes in the rack. We show several time-series snapshots, associated with the three experiments we performed. All metrics interact organically: whenever the predicted CPU temperature's 8th decile crosses the \(T_{hot}^i\) level (i.e., at least 20\% of the CPUs are in a hot state) a decrease in \(T_{rcu}\) ensues, whose speed depends on the gap between the two. Confirming the data in Figure~\ref{deepest:coolinghist}, setting \(T_{hot}^i\) to 68C results in unstable behavior, with \(T_{rcu}\) oscillating between 35C and 45C, while the 78C setting results in a fixed 45C \(T_{rcu}\). 73C, on the other hand, seems to be a sweet spot between the two, with \(T_{rcu}\) staying at 45C most of the time but occasionally dipping down upon high load. Looking at the original CPU temperature data confirmed these behaviors.

\begin{figure}[b]
\centering
\includegraphics[width=0.475\textwidth,trim={10 0 10 0}, clip=true]{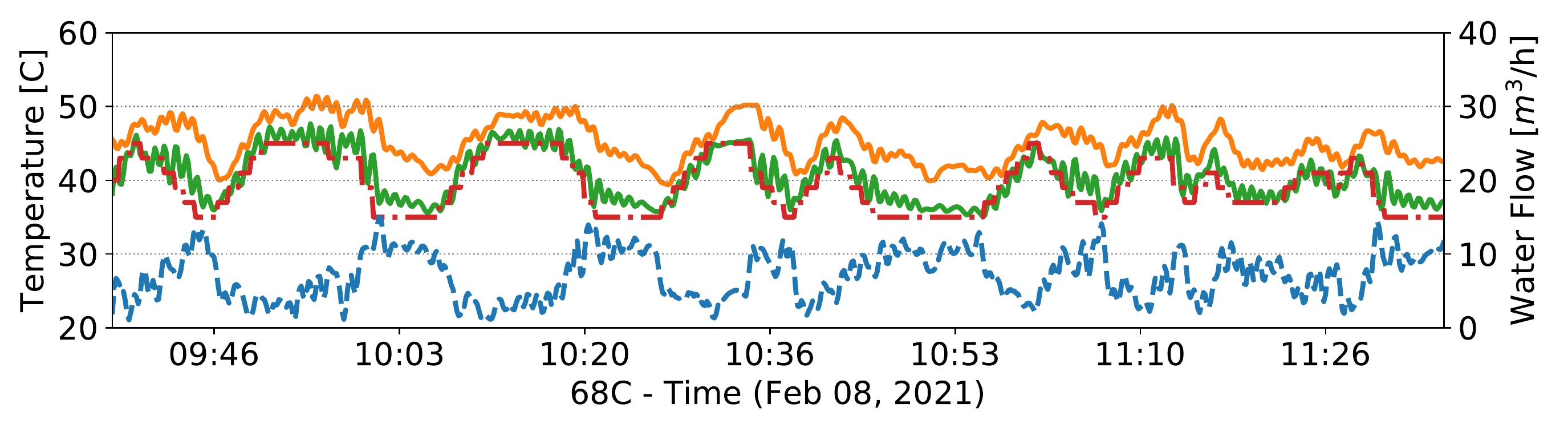} \\
\includegraphics[width=0.475\textwidth,trim={10 0 10 0}, clip=true]{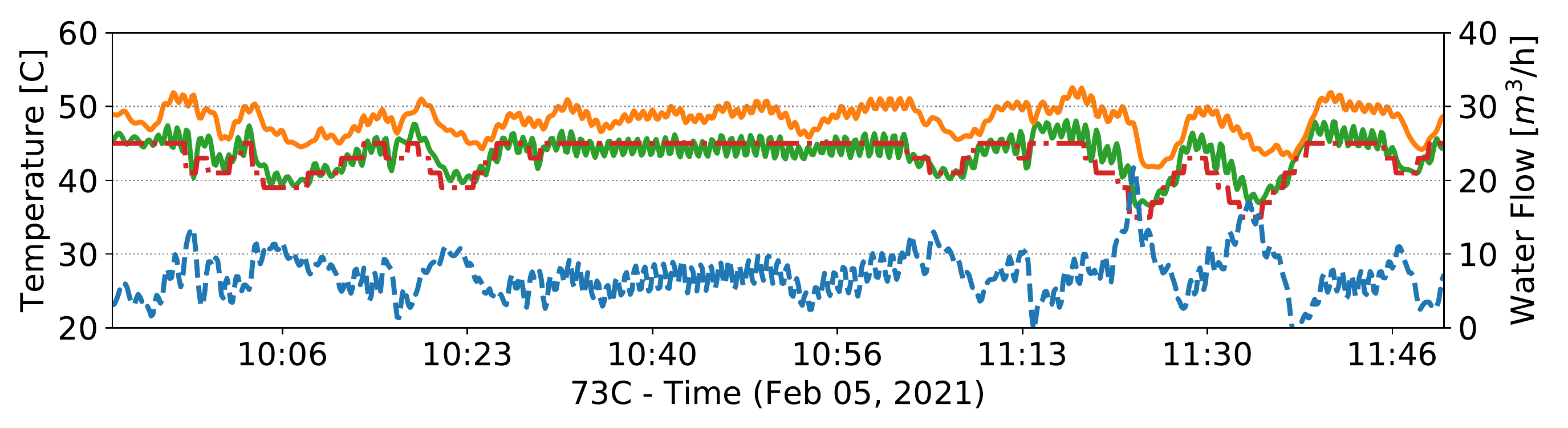} \\
\includegraphics[width=0.475\textwidth,trim={10 0 10 0}, clip=true]{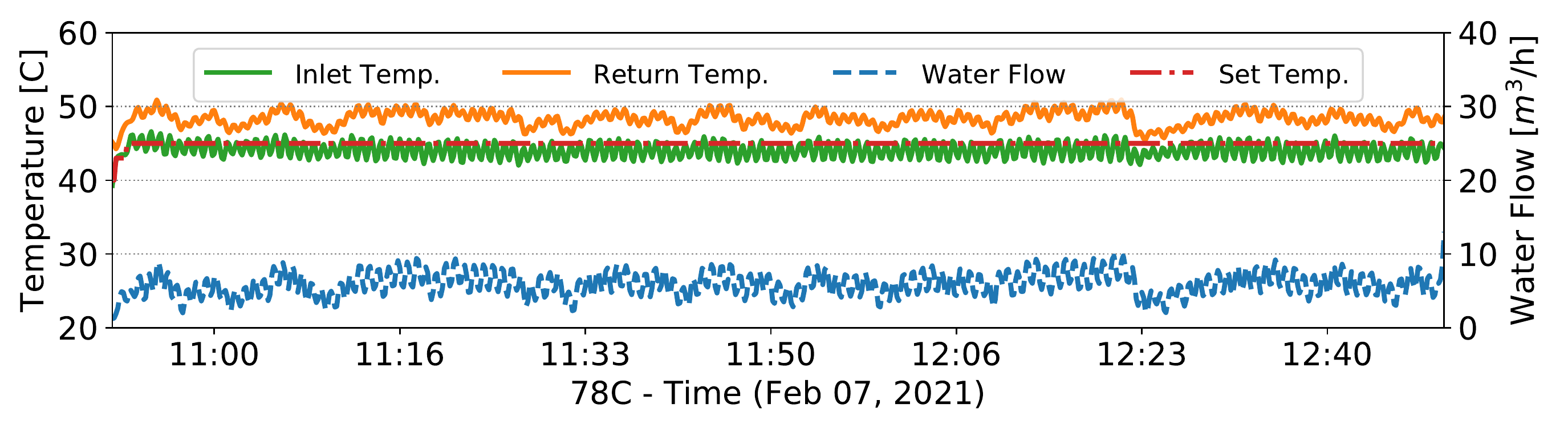}
\caption{Impact of \CM's \(T_{rcu}\) on inlet water flow, as well as on inlet and return temperatures with different \(T_{hot}^i\) values.}
\label{deepest:coolinginf}
\end{figure}

To conclude our analysis, in Figure~\ref{deepest:coolinginf} we show several \CM RCU metrics for the same time frames as in Figure~\ref{deepest:coolingtime}. On top of \(T_{rcu}\), we show the effective secondary inlet and return temperatures and the primary flow rate, which quantifies the amount of cold water coming from the building infrastructure that is consumed per unit of time. Aside from the expected impact of \(T_{rcu}\) on the secondary inlet and return temperatures, there is a substantial effect on the water flow rate: switching between 35C and 45C as \(T_{rcu}\) in a short time, in fact, translates to drastic flow rate differences of up to 10 cubic meters per hour. The temperature difference between secondary inlet and return water lies between 5C and 10C, which is expected from this type of system~\cite{conficoni2015energy, wilde2017coolmuc}.

\subsubsection{Estimation of Benefits}

\begin{figure}[t]
\centering
\captionsetup[subfigure]{}
\subfloat[\CM.]{
\includegraphics[width=0.23\textwidth,trim={0 0 10 0}, clip=true]{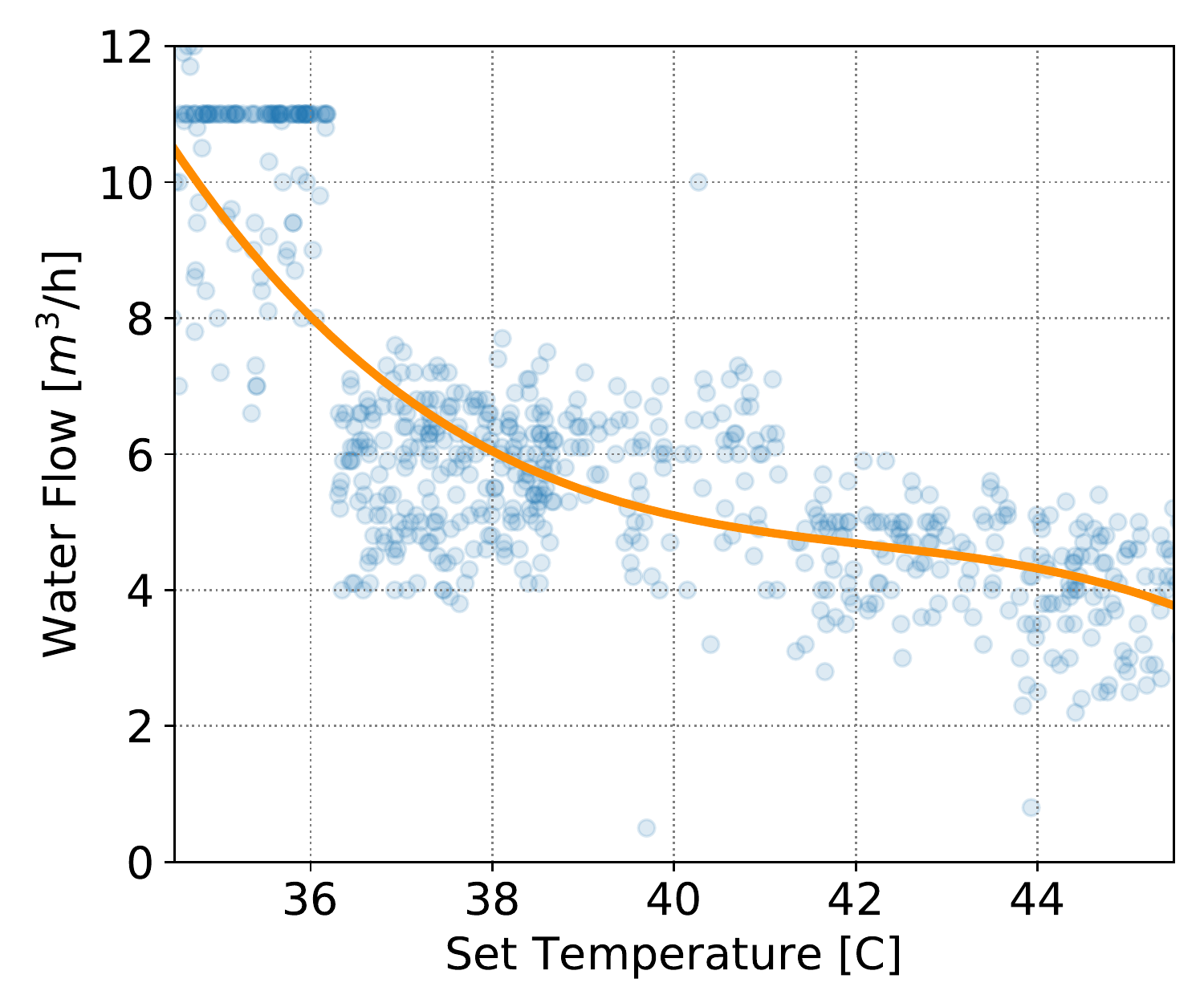}
  }
\subfloat[\ESB.]{
\includegraphics[width=0.23\textwidth,trim={0 0 10 0}, clip=true]{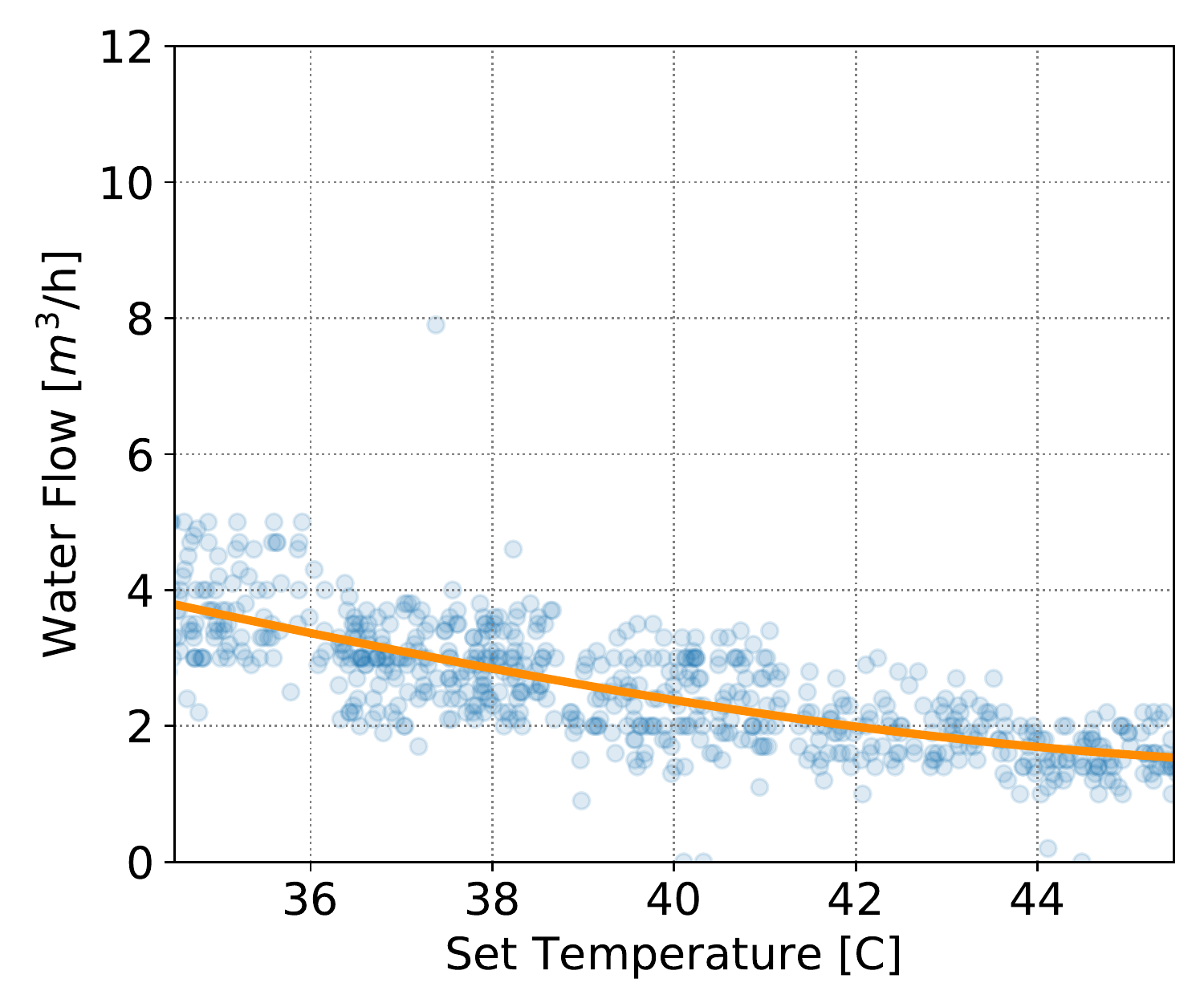}
}
\caption{Fitted inlet water flow rate in function of \(T_{rcu}\) for \CM and \ESB RCUs, computed from observed data.}
\label{deepest:coolingflow}
\end{figure}

Due to the many factors driving a cooling infrastructure, estimating the energy efficiency benefits of our control pipeline is difficult at this point in time - however, we can still derive significant insights. First, adopting a very low \(T_{hot}^i\) setting (e.g., 68C on the \CM) is undesirable due to the large inlet water flow rate fluctuations shown in Figure~\ref{deepest:coolinginf}, which make estimating the demand for cold water under normal operation difficult. In Figure~\ref{deepest:coolingflow} we show the fitted water flow rate in function of \(T_{rcu}\) for the \CM and \ESB racks: in both cases, continuous operation at a \(T_{rcu}\) of 45C leads to a 50\%-lower flow rate compared to 35C - this can be easily obtained with our control pipeline, based on the histograms in Figure~\ref{deepest:coolinghist}. We can thus assume a proportional long-term reduction in energy consumption for pumps and other machinery when scaling our approach to an entire data center, due to the reduced cooling demand. Further, since a high \(T_{rcu}\) enables use of adsorption chilling and \emph{free cooling} (i.e., without any active devices), energy savings may be in reality much more significant.

%% file: sections/LessonsLearned.tex
\section{Lessons Learned and Action Items}
\label{section:lessonslearned}

Sections~\ref{section:sngdeployment} and~\ref{section:dpdeployment} show how our tool chain helps clear the technical challenges of production ODA infrastructures, satisfying in turn the requirements laid out in Section~\ref{section:design}. The \sng deployment shows how \dcdb and \daf are suitable for large-scale HPC systems (\emph{scalability}) and how they can be easily integrated with legacy infrastructures (\emph{flexibility}); our \dpp deployment, on the other hand, shows how many simple \daf plugins can be used for a complex ODAC use case (\emph{modularity}), combining in-band and out-of-band data in the same tool (\emph{holism}). In both cases, as demonstrated also in prior work~\cite{netti2019dcdb, netti2019wintermute, netti2020correlation}, this is achieved with minimal overhead (\emph{footprint}). Still, there are many unaddressed complexity factors behind this process, many of which pertaining the long-term maintainability of ODA and seldom discussed in the literature: we now share our insights about them, and propose a series of associated mitigation actions.

\subsection{Lessons Learned}

ODA comes with a series of complexity factors that are both technical and human: while these relate to ODA in general, ODAC is in practice subject to tighter constraints than ODAV, due its more direct and autonomous impact on operations. 

\paragraph{Allocation of ODA Resources} 

Our optimized \sng deployment requires 2.5TB of Cassandra storage space with 60,000 inserts per second - if we stored all available sensors, with 700,000 inserts per second, storage requirements would easily reach several TBs for each day of operation. Furthermore, both our \sng and \dpp deployments require dedicated hardware resources, with respectively 10 and 3 machines. It is therefore clear that allocating resources for monitoring and ODA requires a significant amount of thought at system design time, with storage being a critical factor: insufficient storage space or bandwidth can result in a very short time-to-live for sensor data, long sampling intervals or less available sensors, severely crippling the effectiveness of ODA.


\paragraph{Orchestration Issues}

On \sng, we discovered soon after deployment that \dcdb Pushers were silently stopping sampling CPU performance counters; this was found to be due to a kernel Perfevent bug, resolved in recent versions of the SLES operating system, which prevented concurrent sampling between \dcdb and the EAR~\cite{corbalan2019ear} framework also used on this system for CPU frequency tuning. This is a common theme in monitoring and ODA: as they are often added during a system's lifetime, they may not have absolute control over data sources and system knobs due to the presence of other frameworks. This aspect can be detrimental if multiple frameworks affect the same knobs, requiring non-trivial orchestration mechanisms for safe operation.


\paragraph{Intrinsic Complexity} 

After deploying \dcdb on the \dpp system, we were alerted by occasional large, unrealistic readings produced by the Intel RAPL interface's CPU energy counters: the associated wrap-around value was found to be the culprit, as it did not comply to the underlying register's width, but to an arbitrary value indicated in a secondary SysFS file. As sensor data was still available, this only led to suspiciously bad performance of the CPU temperature models. Issues of this kind are even harder to spot on large-scale systems such as \sng, exposing 6,8 million sensors: ensuring that all of them are functioning properly is a daunting task~\cite{gimenez2017scrubjay}. From our experience, monitoring data has three levels of consistency: \emph{numerical} (i.e., against a unit of measurement), \emph{spatial} (i.e., across components) and \emph{temporal} (i.e., over time). Consistency issues are hard to diagnose and impair the functionality of ODA models in non-obvious ways.


\paragraph{Benefits Estimation} 

Our cooling control pipeline was deployed for the first time on \dpp, and as such its benefits could only be loosely estimated; as we operated in the context of a research project, this did not hinder us. In general, the importance of monitoring is by now established in data center operations but, as ODA is perceived as an experimental field, use of complex pipelines may encounter significant resistance. Similarly, users may be skeptical of ODA knobs exposed to them (e.g., for CPU frequency tuning), which they might see as a hindrance to performance rather than beneficial to it. Hence, the adoption of ODA should be always proposed with an accompanying quantitative analysis, forecasting its impact with clear metrics such as the \emph{Power Usage Effectiveness} (PUE)~\cite{yuventi2013critical, Bourassa:2019:ODA:3339186.3339210}. Carrying out such an analysis may not be straightforward: the value-add of many ODA techniques is either only proven in theory, or it is clear only after multiple years of operation.


\paragraph{Competence Asymmetry}

Throughout our experiences, we observed that ODA researchers tend to possess a different set of competences compared to system administrators: while the former are usually well-versed in research fields such as data mining, it is the latter that possess the expertise to ensure proper operation of a large-scale platform. This includes knowledge of installation and logging tools, management of permissions and access to infrastructure machines. Bridging the gap between researchers and administrators is thus essential, most notably through coordination. A second issue is the fact that the expertise of most ODA researchers is very specific: satisfactory ODA results are often the product of years of work in highly specific domains, spent training and maintaining models. As it stands, a pervasive use of ODA would require impractical amounts of highly specialized manpower.


\subsection{Action Items}
\label{section:actionitems}

The critical points described above call for efforts to further simplify the adoption of ODA in data centers. We formulate these as \emph{action items} that the ODA research community should undertake.

\paragraph{Portability and Generality}
Making pervasive ODA attainable inevitably passes through the use of open-source and generic tools by the community, coupled with the sharing of competences across institutions. Creating public ODA models that are portable to multiple system architectures is also important: for example, a data center with years of expertise in failure prediction could create and update machine learning models that can be leveraged by other centers, reducing maintenance efforts. \dcdb, \daf and the \csmethod method were designed with this scenario in mind.

\paragraph{Commitment to ODA}
Committing to ODA at the procurement stage of a system is valuable: it allows for defining the appropriate hardware resources, as well as integration efforts with other tools, the associated responsibilities and personnel requirements. In the currently too rare cases where ODA reaches production, this happens at a late stage in a system's life - hence, better management and policing guidelines would provide the ODA community with means to shape future data centers in a more meaningful way.

\paragraph{Maintainability of Models}
ODA models are often susceptible to the changes in operational conditions that occur throughout a system's life: this is especially true for supervised machine learning models, such as the one in Section~\ref{section:dpdeployment}, whose effectiveness depends on the training methodology. In order to improve maintainability, it would be beneficial to employ techniques (e.g., reinforcement learning) that can adapt to changing user workloads, component aging and other operational factors.

\paragraph{Self-Monitoring Mechanisms}
Ensuring the consistency of monitoring data is paramount for ODA models, but this is an excessive burden for system administrators. Hence, automated mechanisms for self-monitoring of ODA pipelines are desirable: these could be as simple as the Health Checker mechanism discussed in Section~\ref{section:dpdeployment}, or could use more complex, statistical approaches to detect deviations and gaps in the monitoring data.

%% file: sections/Conclusions.tex
\section{Conclusions}
\label{section:conclusions}

In this work we presented our ODA experiences in production HPC environments, using the open-source \dcdb and \daf tools, as well as the \csmethod method. We cover a wide range of applications: the \sng job data aggregation deployment is an example of ODAV on large-scale systems, while our \dpp cooling control pipeline is an example of ODAC on a modular, heterogeneous HPC system. Having cleared the challenge of deploying complex ODA pipelines in production, we discuss the main lessons learned from our experiences, driving the push for a pervasive adoption of ODA in data centers. The action items in Section~\ref{section:actionitems} target unresolved issues in the ODA field, ranging from the generality of the models used for most techniques, to the need for a more prominent role of ODA at system procurement, as well as for self-monitoring mechanisms to simplify day-to-day operations. These items, which are directed towards the ODA community as a whole, are meant to ensure that ODA is not only usable on production HPC systems, but also maintainable in the long term. 

On top of the above, we aim to characterize the long-term improvements in energy efficiency and user experience deriving from our ODA deployments. We also plan to extend \daf's ODA capabilities: \dcdb already supports the storage of event data including, for example, application-level instrumentation as well as telemetry about data migrations in hierarchical storage systems. Integrating this data into the \daf workflow will allow for more powerful and precise ODA control.

\vspace{2mm}
\textit{Acknowledgements.} This research activity has received funding from the DEEP-EST project under the EU H2020-FETHPC-01-2016 Programme grant agreement n° 754304. We'd like to thank MEGWARE GmbH for granting us access to infrastructure data and control on the \dpp system.